%% file: 0_main.tex
\PassOptionsToPackage{prologue,table,xcdraw}{xcolor}
\documentclass[manuscript,screen]{acmart}

\usepackage{balance}
\usepackage{enumitem}
\usepackage{bm}
\usepackage{algorithm}
\usepackage{algpseudocode}
\usepackage{multirow} 
\usepackage{amsmath}
\allowdisplaybreaks[4]

\AtBeginDocument{%
  }

\setcopyright{acmlicensed}
\copyrightyear{2018}
\acmYear{2018}
\acmDOI{XXXXXXX.XXXXXXX}
\acmConference[Conference acronym 'XX]{Make sure to enter the correct
  conference title from your rights confirmation email}{June 03--05,
  2018}{Woodstock, NY}
\acmISBN{978-1-4503-XXXX-X/2018/06}

\begin{document}

\title{Brownian Bridge Diffusion for Sequential Recommendation}

\author{Yimeng Bai}
\orcid{0009-0008-8874-9409}
\affiliation{
  \institution{University of Science and Technology of China}
  \city{Hefei}
  \country{China}
}
\email{baiyimeng@mail.ustc.edu.cn}

\author{Yang Zhang}
\authornote{Corresponding author.}
\orcid{0000-0002-7863-5183}
\affiliation{
  \institution{National University of Singapore}
  \city{Singapore}
  \country{Singapore}
}
\email{zyang1580@gmail.com}

\author{Sihao Ding}
\orcid{0000-0003-1796-8504}
\affiliation{
  \institution{ByteDance China}
  \city{Shanghai}
  \country{China}
}
\email{dingsihao@bytedance.com}

\author{Shaohui Ruan}
\orcid{0009-0003-0150-0445}
\affiliation{
  \institution{ByteDance China}
  \city{Shanghai}
  \country{China}
}
\email{ruanshaohui@bytedance.com}

\author{Han Yao}
\orcid{0009-0006-7322-1578}
\affiliation{
  \institution{ByteDance China}
  \city{Shanghai}
  \country{China}
}
\email{yaohan.harvey@bytedance.com}

\author{Danhui Guan}
\orcid{0009-0009-3540-8608}
\affiliation{
  \institution{ByteDance China}
  \city{Shanghai}
  \country{China}
}
\email{guandanhui@bytedance.com}

\author{Fuli Feng}
\orcid{0000-0002-5828-9842}
\affiliation{
  \institution{University of Science and Technology of China}
  \city{Hefei}
  \country{China}
}
\email{fulifeng93@gmail.com}

\author{Tat-Seng Chua}
\orcid{0000-0001-6097-7807}
\affiliation{
  \institution{National University of Singapore}
  \city{Singapore}
  \country{Singapore}
}
\email{dcscts@nus.edu.sg}

\renewcommand{\shortauthors}{Yimeng Bai et al.}

\begin{abstract}
\input{1_abs}
\end{abstract}




\maketitle

\input{2_intro}
\input{3_pre}
\input{4_method}
\input{5_exp}

\input{6_rel}
\input{7_con}

\begin{acks}
This work is supported by the National Natural Science Foundation of China (62272437).
ChatGPT was used solely to assist with language polishing, grammar checking, and improving the clarity of some textual descriptions during the preparation of this manuscript. It was not used to generate the research ideas, methodological design, theoretical derivations, experimental design, result analysis, or conclusions of this work. The authors reviewed, edited, and verified all AI-assisted content and take full responsibility for the final content of the paper.
\end{acks}

\bibliographystyle{ACM-Reference-Format}
\balance
\bibliography{9_ref}

\input{8_appd}

\end{document}

%% file: 1_abs.tex
Diffusion models, known for their strong generative capability derived from iterative noising and denoising processes, have recently emerged as a promising paradigm for sequential recommendation. To incorporate user history for personalization, existing methods typically follow a history-guided denoising paradigm inspired by text-guided image generation, where target item representations are reconstructed from Gaussian noise conditioned on user historical interactions. However, this design remains fundamentally anchored to an ``item $\leftrightarrow$ noise'' formulation, introducing an additional noise-reconstruction burden that may distract the model from capturing user-specific preference structures. Motivated by this limitation, we revisit diffusion-based sequential recommendation from a preference-centric perspective and adopt a preference bridging design that enables a direct ``item $\leftrightarrow$ history'' transition instead of relying on Gaussian noise. Based on this idea, we propose \textit{Brownian Bridge Diffusion Recommendation (BBDRec)}, which leverages the Brownian bridge process to construct a structured diffusion trajectory between target items and user historical representations, thereby better aligning diffusion modeling with the intrinsic nature of recommendation. Extensive experiments on multiple public datasets show that BBDRec consistently outperforms representative sequential and diffusion-based recommendation baselines. The implementation code is publicly available at \url{https://github.com/baiyimeng/BBDRec}.

%% file: 2_intro.tex
\section{Introduction}
\begin{figure}
\centering
\includegraphics[width=1.0\textwidth]{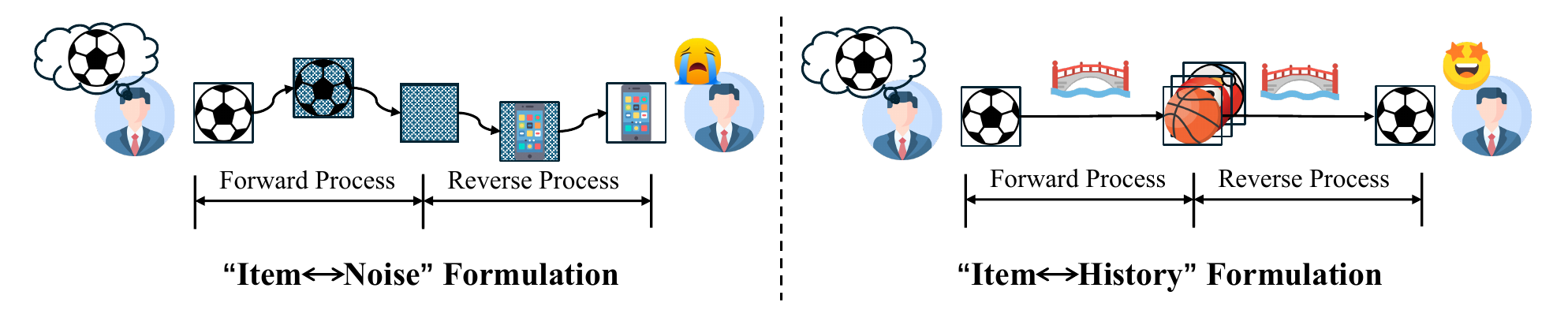}
\caption{Comparison between the conventional noise-anchored diffusion process that models the ``item $\leftrightarrow$ noise'' transition and our preference-centric diffusion process that models the ``item $\leftrightarrow$ history'' transition.}
\Description{The figure compares two diffusion processes for sequential recommendation. The conventional process corrupts a target item representation into Gaussian noise and reconstructs it with user history as conditional guidance. The proposed process directly bridges the target item representation and the user history representation.}
\label{fig:1_compare}
\end{figure}

Inspired by the remarkable success of generative AI, transitioning recommendation paradigms from discriminative to generative approaches has attracted significant attention from academia and industry~\cite{GR, BIGRec, LLM4GR, li2024survey}. Among these efforts, diffusion-based methods~\cite{DreamRec, DiffRec} represent a prominent direction. These methods model the degradation of data and its reverse process to reconstruct the original data, which enables flexible and expressive generative modeling. Recently, Kuaishou\footnote{\url{https://www.kuaishou.com}}, a large-scale short video-sharing platform, implemented a diffusion-based method in its industrial recommender system~\cite{DimeRec}, achieving notable improvements in user satisfaction.

For recommendation tasks, modeling users' historical interactions is crucial. To capture such information, existing diffusion-based recommender systems typically adopt a history-guided item generation paradigm~\cite{DreamRec, DiffuRec, DimeRec, CaDiRec, DiffRIS, iDreamRec, SdifRec, CDDRec}, inspired by text-guided image generation methods~\cite{LDM}. Specifically, as illustrated in Figure~\ref{fig:1_compare}, these methods progressively corrupt target item representations into Gaussian noise during the forward diffusion process. In the reverse denoising process, the model leverages users' historical interactions as conditional guidance to reconstruct the target item representations. This conditional item embedding generation mechanism enables the model to effectively exploit user history information, thereby facilitating personalized recommendations.

However, this noise-anchored diffusion formulation is not well aligned with the characteristics of recommendation tasks and can lead to suboptimal preference modeling. Existing methods require the model to simultaneously learn both the transition between item representations and Gaussian noise distributions and the personalized relationship between user history and target items. As a result, the modeling process tends to emphasize noise reconstruction, while the semantic relationship between historical interactions and target items is only implicitly captured through conditional guidance~\cite{DDBM, SCFG, guidance_scale, NoiseRefine}. Although such a formulation is effective in text-to-image generation, where diffusion models help bridge the modality gap between textual conditions and image representations~\cite{LDM}, recommendation scenarios differ substantially because user history and target items are inherently represented within the same semantic embedding space. Therefore, introducing Gaussian noise as an intermediate state may not be the most suitable choice and can increase the learning difficulty of preference modeling~\cite{BBDM}.

Motivated by these observations, we revisit diffusion modeling for sequential recommendation from a preference-centric perspective. Rather than following the conventional ``item $\leftrightarrow$ noise'' modeling paradigm, we designate the user history representation as the terminal state of the forward diffusion process and, correspondingly, as the initial state of the reverse process. Under this formulation, the forward process can be interpreted as gradually corrupting a specific target item into an ambiguous preference state implicitly encoded by the user history, while the reverse process decodes the target item from this history-conditioned preference state. In contrast to the noise-anchored diffusion, this preference bridging design focuses entirely on modeling the transition between the target item and the history representation, thereby enabling more effective exploitation of sequential behavioral information.

To model the ``item $\leftrightarrow$ history'' relationship, the key is to construct a noising process that follows a constrained trajectory toward a specific semantic endpoint, rather than diffusing representations toward an unstructured Gaussian prior as in conventional diffusion models. To this end, we propose a novel framework, termed \textit{Brownian Bridge Diffusion Recommendation} (BBDRec). The core idea is to leverage the Brownian bridge process~\cite{BB, BBDM, SBBB} to enforce an endpoint-constrained stochastic trajectory. Instead of adopting a purely noise-anchored diffusion process, BBDRec guides the forward transition through a time-varying bridge between the target item representation and the user history representation, gradually shifting the target-side representation toward the history-side representation. Additionally, we propose a customized progressive learning strategy that first establishes a preference-aware representation space and then initializes BBDRec with the learned embeddings for diffusion-based fine-tuning, thereby coordinating diffusion learning and recommendation modeling at the representation level. Empirical evaluations on multiple public datasets consistently demonstrate its superior performance in sequential recommendation tasks.

The main contributions of this work are summarized as follows:
\begin{itemize}[leftmargin=*]
\item We reformulate diffusion-based sequential recommendation as a preference bridging design between user history representations and target items, moving beyond the conventional noise-anchored diffusion paradigm.

\item We propose BBDRec to instantiate the modeling of the ``item $\leftrightarrow$ history'' transition, together with a customized progressive learning strategy that coordinates diffusion learning and recommendation modeling.

\item We conduct extensive experiments on multiple public datasets, showing that BBDRec consistently outperforms representative sequential and diffusion-based recommendation baselines.
\end{itemize}

%% file: 3_pre.tex
\section{Preliminary}\label{sec:pre}

Let $\mathcal{D}$ represent the collected user-item interaction data. We denote a sample in  $\mathcal{D}$ by $(s, y) \in \mathcal{D}$, where $s$ denotes a user's historical interaction sequence before the next interaction, and $y$ denotes the next item interacted with by the user. 
{Notably, each user may have multiple samples in $\mathcal{D}$, with different interactions treated as the next item $y$.}
Our objective is to train a diffusion-based recommender based on $\mathcal{D}$, which can recommend suitable items from the entire item pool for the user's next interaction in a generative manner, given users' historical interaction sequences. 
Before introducing our method, we first outline the current noise-anchored diffusion recommendation framework that underpins this work, which comprises the following two critical processes:

\vspace{+3pt}
\textbf{Forward Process (Item $\rightarrow$ Noise).} 
This process progressively perturbs the target item of each sample, \textit{i.e.}, $y$ of $(s,y)$,  by introducing Gaussian noise over multiple time steps. Specifically, for each sample $(s, y)$, the embedding of the target item $y$ is treated as the initial clean representation and progressively perturbed. Let $\bm{x}_0$ represent the starting point of the process, having $\bm{x}_0 = \bm{e}_y$. Then, $\bm{x}_0$ is gradually perturbed across multiple steps with noising levels increasing according to a predefined schedule $[\alpha_1, \ldots, \alpha_T]$, where $T$ is the maximum diffusion step. Formally, the transition process at each step is defined as: 
\begin{equation}\label{eq:DDPM-forward}
    q(\bm{x}_t|\bm{x}_{t-1})=\mathcal{N}(\bm{x}_t;\sqrt{\alpha_t}\bm{x}_{t-1},(1-\alpha_t) \bm{I}),
\end{equation}
where $\bm{x}_t$ denotes the noisy representation at step $t$, and $\alpha_t \in [0,1]$ controls the amount of signal retained from the previous step. 
By defining $\bar\alpha_t=\prod_{i=1}^{t}\alpha_i$, $\bm{x}_t$ can also be efficiently obtained through a multi-step transition:
\begin{equation}\label{eq:DDPM-multistep}
    q(\bm{x}_t|\bm{x}_0)=\mathcal{N}(\bm{x}_t;\sqrt{\bar\alpha_t}\bm{x}_{0},(1-\bar\alpha_t) \bm{I}).
\end{equation}
When $T$ is sufficiently large and $\bar{\alpha}_T$ approaches zero, the endpoint $\bm{x}_T$ can be approximately regarded as Gaussian noise. In the forward process, no model parameters need to be optimized.

\vspace{+3pt}
\textbf{Reverse Process (Noise $\rightarrow$ Item)}. 
The reverse process gradually reconstructs the clean target item embedding from $\bm{x}_T$, which is approximately distributed as Gaussian noise, conditioned on the user history $s$. 
Given the noisy representation at step $t$ (\emph{i.e.}, $\bm{x}_t$) and the history $s$, the representation at step $t-1$ ({\emph{i.e., }} $\bm{x}_{t-1}$) is obtained through  the following denoising process:
\begin{equation}\label{eq:DDPM-reverse}
p_\theta(\bm{x}_{t-1}|\bm{x}_t, s) 
= \mathcal{N}(\bm{x}_{t-1}; {\bm{\hat\mu}}_\theta(\bm{x}_t, t, s), \bm{\Sigma}_\theta(\bm{x}_t,t,s)), 
\end{equation}
where ${\bm{\hat\mu}}_\theta(\bm{x}_t, t, s)$ and $\bm{\Sigma}_\theta(\bm{x}_t,t,s)$ are the Gaussian parameters produced by a neural network with learnable parameters $\theta$.
In the reverse process, the parameter $\theta$ needs to be learned. Its optimization is carried out by maximizing the Evidence Lower Bound (ELBO), which is approximated by minimizing~\cite{Understanding}:
\begin{equation}
D_{\mathrm{KL}}\left(q(\bm{x}_{t-1}|\bm{x}_t, \bm{x}_0) \Vert p_\theta(\bm{x}_{t-1}|\bm{x}_t, s)\right),
\end{equation}
where $q(\bm{x}_{t-1}|\bm{x}_t, \bm{x}_0)$ denotes the posterior distribution, which has the following closed form:
\begin{equation}\label{eq:DDPM-post}
\begin{aligned}
&q(\bm{x}_{t-1}|\bm{x}_t, \bm{x}_0) 
= \mathcal{N}(\bm{x}_{t-1}; {\bm{\mu}}_t(\bm{x}_t, \bm{x}_0), {\sigma}_t^2 \bm{I}),  \\
&{\bm{\mu}}_t(\bm{x}_t, \bm{x}_0) 
= \frac{\sqrt{\alpha_t}(1-\bar{\alpha}_{t-1})}{1-\bar{\alpha}_t}\bm{x}_t 
+ \frac{\sqrt{\bar{\alpha}_{t-1}}(1-\alpha_t)}{1-\bar{\alpha}_t}\bm{x}_0, \\
&{\sigma}_t^2 
= \frac{(1-\alpha_t)(1-\bar{\alpha}_{t-1})}{1-\bar{\alpha}_t}.
\end{aligned} 
\end{equation}
To ensure training stability and reduce computational complexity, $\bm{\Sigma}_\theta(\bm{x}_t,t,s)$ in Equation~\eqref{eq:DDPM-reverse} is set as $\sigma_t^2\bm{I}$~\cite{DDPM,DiffRec}. 
This can simplify the optimization of the KL divergence to aligning $\bm{\hat\mu}_\theta(\bm{x}_t, t, s)$ with $\bm{\mu}_t(\bm{x}_t, \bm{x}_0)$.
Building on this and following the definition of $\bm{\mu}_t(\bm{x}_t, \bm{x}_0)$ in Equation~\eqref{eq:DDPM-post},
$\hat{\bm{\mu}}_\theta(\bm{x}_t, t, s)$ in Equation~\eqref{eq:DDPM-reverse} can be obtained using the reparameterization technique as follows:
\begin{equation}\label{eq:DDPM-reparam}
\hat{\bm{\mu}}_\theta(\bm{x}_t, t, s) 
= \frac{\sqrt{\alpha_t}(1-\bar{\alpha}_{t-1})}{1-\bar{\alpha}_t}\bm{x}_t 
+ \frac{\sqrt{\bar{\alpha}_{t-1}}(1-\alpha_t)}{1-\bar{\alpha}_t}f_\theta(\bm{x}_t,t,s),
\end{equation}
where $f_\theta(\cdot)$ represents a denoiser model used to predict $\bm{x}_0$ (with $\bm{x}_0=\bm{e}_y$)  based on the state at each diffusion step. Finally, 
we can perform optimization by simply minimizing the MSE loss between $\bm{x}_0$ and the model output $f_\theta(\bm{x}_{t}, t, s)$:
\begin{equation}\label{eq:DDPM-loss}
\mathcal{L}_\text{denoise}
=\sum_{(s,y)\in\mathcal{D}}\mathbb{E}_{t} [\Vert \bm{e}_y-f_\theta(\bm{x}_t,t,s)\Vert_2^2].
\end{equation}
Notably, this history-guided design is typically realized by explicitly incorporating the user history $s$ into the denoiser model $f_\theta(\bm{x}_t, t, s)$, so that the reverse diffusion process is indirectly guided toward reconstructing item representations consistent with user preferences.

%% file: 4_method.tex
\section{Methodology}

\begin{figure}[t]
\centering
\includegraphics[width=1.0\textwidth]{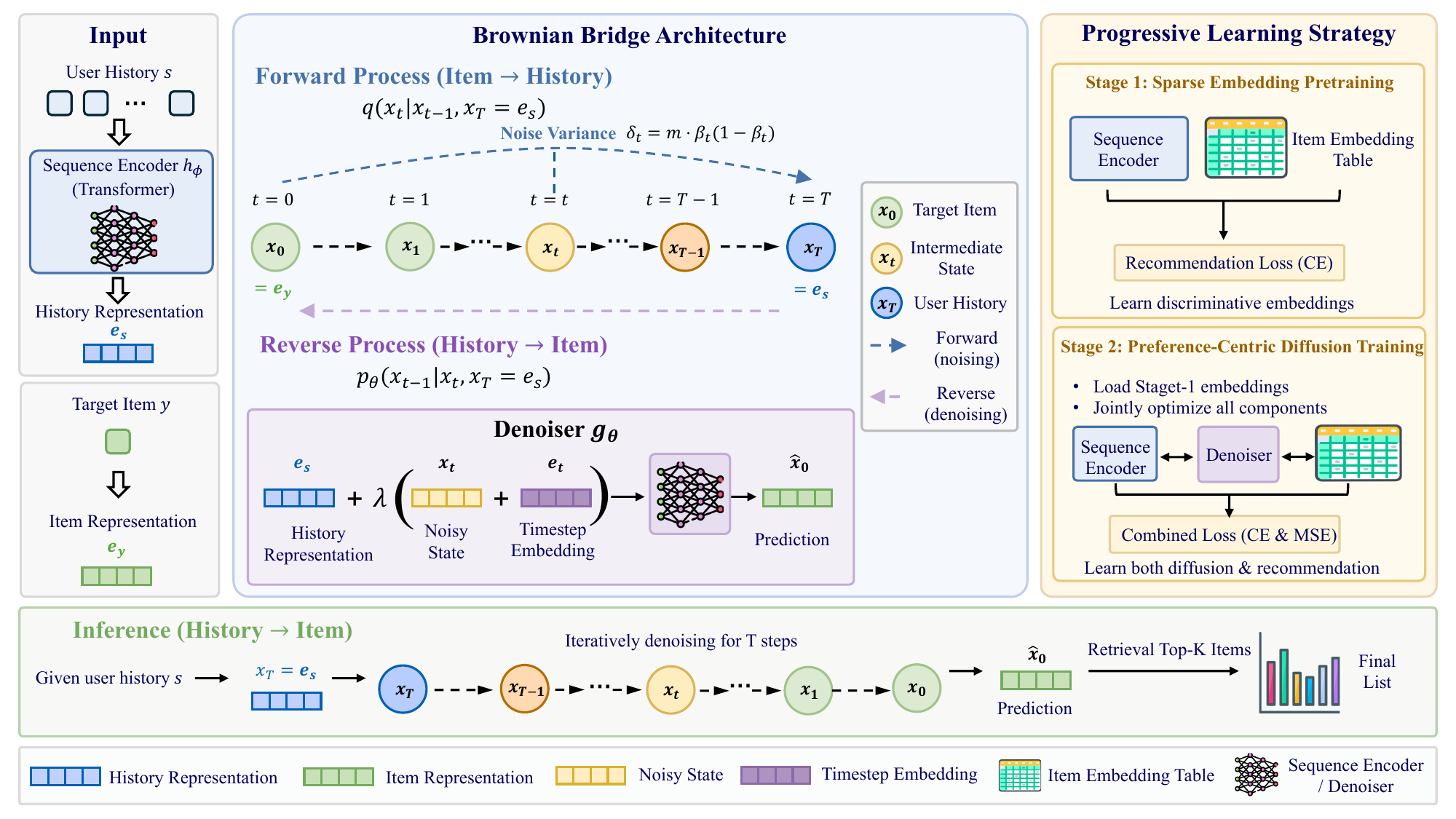}
\caption{An overview of the proposed BBDRec framework. The Brownian bridge architecture enables a direct transition between target item representations and sequential history representations through a structured bridge diffusion process. The progressive learning strategy coordinates diffusion learning and recommendation modeling through representation pretraining and diffusion-based fine-tuning.}
\label{fig:frame}
\Description{The figure illustrates the BBDRec framework, including a Brownian bridge diffusion process between target item representations and user history representations, a denoiser for reconstructing target item representations, and a progressive learning strategy with embedding pretraining and diffusion-based fine-tuning.}
\end{figure}

We revisit diffusion modeling for sequential recommendation from a preference-centric perspective. Rather than modeling the conventional ``item $\leftrightarrow$ noise'' transition, we reformulate diffusion as a transition process between target item representations and user history representations. The forward process progressively evolves item representations toward the preference state represented by user history, while the reverse process reconstructs target items from this state. In this way, diffusion is naturally aligned with the ``item $\leftrightarrow$ history'' relationship, enabling more effective utilization of user historical behavioral signals.

To realize this preference-bridging design, the key challenge is to construct a structured noising mechanism that progressively evolves toward a specific preference state rather than unconstrained Gaussian noise. To this end, we leverage the Brownian bridge process~\cite{BBDM} and propose the \textit{Brownian Bridge Diffusion Recommendation} (BBDRec) framework. We first provide an overview of the framework, followed by detailed descriptions of the Brownian bridge architecture, the progressive learning strategy, and the training and inference procedures.

\subsection{Overview of BBDRec}

Figure~\ref{fig:frame} provides an overview of BBDRec, which consists of a Brownian bridge architecture for preference-centric diffusion modeling and a progressive learning strategy for effective optimization.

\vspace{+3pt}
\textbf{Brownian Bridge Architecture}. 
The architecture provides a structured noising and denoising mechanism to realize the proposed bridge diffusion process. Specifically, it incorporates a Brownian-bridge-based noising process that progressively steers the diffusion trajectory toward a specific preference state represented by the user history (\emph{i.e.},  $\bm{x}_{T} = \bm{e}_s$). At each diffusion step $t$, rather than solely injecting Gaussian noise as in Equation~\eqref{eq:DDPM-forward}, the transition process gradually incorporates the preference-state information by making the transition function depend on $\bm{x}_{T}$, thereby progressively guiding the diffusion trajectory toward the desired preference representation.

\vspace{+3pt}
\textbf{Progressive Learning Strategy}. 
Directly training the sequence encoder and diffusion model in an end-to-end manner may lead to embedding collapse, as the diffusion objective can over-smooth item representations. To address this issue, we adopt a two-stage training framework. In the first stage, the model is trained with a recommendation objective to learn preference-aware representations. In the second stage, the pretrained parameters are used to initialize BBDRec for subsequent diffusion-based refinement.

\subsection{Brownian Bridge Architecture}\label{sec:BBA}

Overall, the architecture is designed to directly model the transition between item representations and user history representations based on the Brownian bridge process~\cite{BBDM}. To evolve $\bm{x}_0$ toward a specific preference state $\bm{x}_T$ instead of unconstrained Gaussian noise, the Brownian bridge process progressively increases the influence of the endpoint $\bm{x}_T$ on intermediate states during diffusion, which can be formulated as:
\begin{equation}\label{eq:BB}
\begin{aligned}
&q(\bm{x}_t|\bm{x}_0,\bm{x}_T)=\mathcal{N}(\bm{x}_t;(1-\beta_t)\bm{x}_0+\beta_t\bm{x}_T,\delta_t\bm{I}),\\
& \beta_t=\frac{t}{T}, \; \quad \delta_t=m\cdot \beta_t(1-\beta_t).
\end{aligned}
\end{equation}
At $t=0$, the distribution degenerates to $\bm{x}_0$, while at $t=T$, it degenerates to $\bm{x}_T$, indicating that the diffusion trajectory converges to the designated preference state. As $t$ increases from $0$ to $T$, the variance first increases and then decreases, reaching its maximum value at $t=T/2$, corresponding to the highest uncertainty state during diffusion. Therefore, the diffusion intensity can be controlled through the hyper-parameter $m$.
Building on Equation~\eqref{eq:BB}, we develop our preference-centric diffusion process by setting $\bm{x}_{0}=\bm{e}_{y}$ (target item representation) and $\bm{x}_{T}=\bm{e}_s$ (user history representation).

\vspace{+3pt}
\textbf{Forward Process (Item $\rightarrow$ History)}.
In the forward process, the diffusion trajectory progressively evolves from the target item representation toward the preference state represented by the user history. Specifically, the transition probability at each step is conditioned on the previous state $\bm{x}_{t-1}$ and the preference state $\bm{x}_{T}=\bm{e}_s$, where $\bm{e}_s=h_\phi(s)$ is obtained by a Transformer-based sequence encoder parameterized by $\phi$, a common setting in sequential recommendation. The transition probability can be derived as\footnote{Proofs for this and subsequent results are provided in Appendix.}:
\begin{equation}\label{eq:BB-forward}
\begin{aligned}
& q(\bm{x}_t|\bm{x}_{t-1},\bm{x}_T)=\mathcal{N}(\bm{x}_t;\gamma_t\bm{x}_{t-1}+(\beta_t-\gamma_t\beta_{t-1})\bm{x}_T, \hat{\delta}_{t}\bm{I}),\\
& \gamma_t=\frac{1-\beta_t}{1-\beta_{t-1}},\;\hat{\delta}_{t}=\delta_t-\gamma_t^2\delta_{t-1},
\end{aligned}
\end{equation}
where $\beta_{t}$ and $\delta_{t}$ are defined in Equation~\eqref{eq:BB}.

\vspace{+3pt}
\textbf{Reverse Process (History $\rightarrow$ Item)}.
In the reverse process, the objective is to progressively reconstruct the target item representation $\bm{x}_{0}=\bm{e}_{y}$ from the preference state $\bm{x}_{T} = \bm{e}_{s}$. According to the Brownian bridge formulation, each intermediate state $\bm{x}_{t-1}$ is generated based on $\bm{x}_{t}$ and the preference state $\bm{x}_{T}$ as follows:
\begin{equation}\label{eq:BB-post}
\begin{aligned}
p_\theta(\bm{x}_{t-1}|\bm{x}_t,\bm{x}_T)=\mathcal{N}(\bm{x}_{t-1};\hat{\bm{\mu}}_\theta(\bm{x}_t,t,\bm{x}_T),\Sigma_\theta(\bm{x}_t,t,\bm{x}_T)),
\end{aligned}
\end{equation}
where $\hat{\bm{\mu}}_\theta(\bm{x}_t,t,\bm{x}_T)$ and $\Sigma_\theta(\bm{x}_t,t,\bm{x}_T)$ are generated by a neural network parameterized by $\theta$, given $\bm{x}_t$, $\bm{x}_{T}$, and the diffusion step $t$.
Following Section~\ref{sec:pre}, the model parameters are optimized by maximizing the ELBO, which can be approximated by minimizing the KL divergence:
\begin{equation}
D_{\mathrm{KL}}\left(
q(\bm{x}_{t-1}|\bm{x}_t,\bm{x}_0,\bm{x}_T)
\Vert
p_\theta(\bm{x}_{t-1}|\bm{x}_t,\bm{x}_T)
\right),
\end{equation}
where the posterior distribution is given by:
\begin{equation}\label{eq:BB-reverse}
\begin{aligned}
& q(\bm{x}_{t-1}|\bm{x}_t,\bm{x}_0,\bm{x}_T)=\mathcal{N}(\bm{x}_{t-1};{\bm{\mu}}_t(\bm{x}_t,\bm{x}_0,\bm{x}_T),\tilde{\delta}_t\bm{I}), \\
& {\bm{\mu}}_t(\bm{x}_t,\bm{x}_0,\bm{x}_T) = \frac{\delta_{t-1}}{\delta_{t}}\gamma_t\bm{x}_t+\frac{\hat{\delta}_{t}}{\delta_{t}}(1-\beta_{t-1})\bm{x}_0 +(\beta_{t-1}-\frac{\delta_{t-1}}{\delta_{t}}\gamma_t\beta_t)\bm{x}_T,\quad\tilde{\delta}_t=\frac{\hat{\delta}_{t}\delta_{t-1}}{\delta_{t}}.
\end{aligned}
\end{equation}
By setting $\Sigma_\theta(\bm{x}_t, t, \bm{x}_T)=\tilde{\delta}_t \bm{I}$, the optimization objective reduces to aligning the predicted mean $\hat{\bm{\mu}}_\theta(\bm{x}_t,t,\bm{x}_T)$ with the posterior mean ${\bm{\mu}}_t(\bm{x}_t,\bm{x}_0,\bm{x}_T)$.
Using the reparameterization formulation, the predicted mean can be expressed as:
\begin{equation}\label{eq:BB-reparam}
\begin{aligned}
& \hat{\bm{\mu}}_\theta(\bm{x}_t,t,\bm{x}_T) = \frac{\delta_{t-1}}{\delta_{t}}\gamma_t\bm{x}_t+\frac{\hat{\delta}_{t}}{\delta_{t}}(1-\beta_{t-1})g_\theta(\bm{x}_t,t,\bm{x}_T) +(\beta_{t-1}-\frac{\delta_{t-1}}{\delta_{t}}\gamma_t\beta_t)\bm{x}_T,
\end{aligned}
\end{equation}
where $g_\theta(\cdot)$ denotes a denoiser model for predicting $\bm{x}_0$. Consequently, the objective can be simplified to aligning $g_\theta(\bm{x}_t,t,\bm{x}_T)$ with $\bm{x}_0$, which is achieved by minimizing the following MSE loss:
\begin{equation}\label{eq:BB-loss-diff}
\mathcal{L}_\text{diff} = \sum_{(s, y) \in \mathcal{D}} \mathbb{E}_t \Vert \bm{e}_y - g_\theta(\bm{x}_t,t,\bm{x}_T) \Vert_2^2.
\end{equation}
Following previous work~\cite{DiffuRec}, we implement the denoiser by aggregating different input components and feeding the resulting representation into the model for prediction, which can be formulated as:
\begin{equation}\label{eq:agg}
g_\theta(\bm{x}_t,t,\bm{x}_T)
=
g_\theta\left(\bm{e}_s + \lambda(\bm{x}_t+\bm{e}_t)\right),
\end{equation}
where $\lambda$ is a hyperparameter and $\bm{e}_t$ denotes the timestep embedding. The denoiser $g_\theta(\cdot)$ can be instantiated with either an MLP or a Transformer-based architecture.

\subsection{Progressive Learning Strategy}\label{sec:PLS}

In the overall framework, we need to jointly learn three types of parameters: (i) sparse parameters, \emph{i.e.}, the embedding tables for all items, (ii) dense parameters of the sequence encoder that produces user preference representations, and (iii) dense parameters of the denoiser for preference-centric diffusion modeling. In our experiments, we observe that directly optimizing all components from scratch in an end-to-end manner leads to representation collapse~\cite{embedding_collapse,PreferDiff}.We conjecture that this arises from an inherent tension between diffusion learning, which tends to smooth representation distributions through stochastic perturbations~\cite{PreferDiff}, and recommendation modeling, which requires discriminative and behavior-aware structures in the embedding space. To address this issue, inspired by the optimization approach in LLM-based recommendation~\cite{CoLLM}, we adopt a progressive learning strategy that decomposes the optimization into:

\vspace{+3pt}
\textbf{Stage 1: Sparse Embedding Pretraining.} We first train the sequence encoder together with the embedding tables using a standard recommendation objective, in order to obtain stable and discriminative item representations. The optimization objective is defined as:
\begin{equation}
\mathcal{L}_{1} = - \sum_{(s,y)\in\mathcal{D}} \log \frac{\exp(\bm{e}_s \cdot \bm{e}_y)}{\sum_{y'} \exp(\bm{e}_s \cdot \bm{e}_{y'})},
\end{equation}
where $y'$ ranges over all items in the corpus.

\vspace{+3pt}
\textbf{Stage 2: Preference-Centric Diffusion Training.} 
We then initialize the diffusion framework with the pretrained embeddings learned in Stage 1, providing high-quality representations for the sequence encoder and thereby improving training stability. In addition, we adopt a learning rate warmup strategy for sparse embedding optimization by using a smaller learning rate in the early stage, alleviating disturbances to the embedding space during diffusion training~\cite{multi-epoch}. The overall training objective to jointly optimize all components is formulated as:
\begin{equation}\label{eq:2_rec}
\begin{aligned}
\mathcal{L}_{2} &= \mathcal{L}_{\text{rec}}+\eta \mathcal{L}_{\text{diff}}, \\
\mathcal{L}_{\text{rec}} &= - \sum_{(s,y)\in\mathcal{D}} \log \frac{\exp(g_\theta(\bm{x}_t,t,\bm{x}_T) \cdot \bm{e}_y)}{\sum_{y'} \exp(g_\theta(\bm{x}_t,t,\bm{x}_T) \cdot \bm{e}_{y'})},
\end{aligned}
\end{equation}
where $\mathcal{L}_{\text{diff}}$ is the diffusion objective defined in Section~\ref{sec:BBA}, $\mathcal{L}_{\text{rec}}$ is the recommendation objective, and $\eta$ is a balancing hyperparameter. Unlike prior methods that rely solely on denoising loss or replace it entirely with recommendation loss~\cite{DiffuRec,DreamRec}, we combine both objectives to jointly supervise recommendation learning and diffusion modeling, aligning distribution learning with the recommendation task while preserving the generative capability of diffusion.

Finally, we present the training and inference details of BBDRec, illustrated in Algorithm~\ref{alg:training} and Algorithm~\ref{alg:inference}, respectively. Following~\cite{BIGRec}, the final predicted embedding is grounded into the item space by retrieving the top-$K$ items with the largest inner-product similarities.

\begin{algorithm}[t]
\caption{Training procedure of BBDRec.}
\label{alg:training}
\begin{algorithmic}[1]

\State \textbf{// Stage 1: Sparse Embedding Pretraining}
\While{not converged}
\State $(s,y)\sim\mathcal{D}$\Comment{Sample data from training set}
\State $\bm{e}_s = h_\phi(s)$\Comment{Encode user history}
\State $\mathcal{L}_1 = - \log \frac{\exp(\bm{e}_s \cdot \bm{e}_y)}{\sum_{y'} \exp(\bm{e}_s \cdot \bm{e}_{y'})}$\Comment{Recommendation loss}
\State Update $\phi$ and embedding tables using $\nabla \mathcal{L}_{\text{1}}$
\EndWhile

\State \textbf{// Stage 2: Preference-Centric Diffusion Training}
\State Initialize with learned embeddings from Stage 1

\While{not converged}
\State $(s,y)\sim\mathcal{D}$\Comment{Sample data from training set}
\State $\bm{e}_s = h_\phi(s)$\Comment{Encode user history}
\State $\bm{x}_T=\bm{e}_s,\;\bm{x}_0=\bm{e}_y$\Comment{Preference and item representations}
\State $t\sim \text{Uniform}({1,\dots,T})$\Comment{Sample diffusion step}
\State $\bm{\epsilon}\sim\mathcal{N}(\bm{0},\bm{I})$\Comment{Sample Gaussian noise}
\State $\bm{x}_t = (1-\beta_t)\bm{x}_0+\beta_t\bm{x}_T+\sqrt{\delta_t}\bm{\epsilon}$\Comment{Forward process}
\State $\mathcal{L}_{\text{diff}} = \|\bm{e}_y - g_\theta(\bm{x}_t,t,\bm{x}_T)\|_2^2$\Comment{Diffusion loss}
\State $\mathcal{L}_{\text{rec}} = - \log \frac{\exp(g_\theta(\bm{x}_t,t,\bm{x}_T)\cdot \bm{e}_y)}{\sum_{y'} \exp(g_\theta(\bm{x}_t,t,\bm{x}_T)\cdot \bm{e}_{y'})}$\Comment{Recommendation loss}
\State $\mathcal{L}_2 = \mathcal{L}_{\text{rec}} + \eta\mathcal{L}_{\text{diff}}$\Comment{Total loss}
\State Update $\theta$, $\phi$, and embeddings
\EndWhile

\end{algorithmic}
\end{algorithm}

\begin{algorithm}[t]
\caption{Inference procedure of BBDRec.}
\label{alg:inference}
\begin{algorithmic}[1]
\State $s\sim\mathcal{D}_{\mathrm{test}}
$\Comment{Sample data from testing set}
\State $\hat{\bm{x}}_T=\bm{e}_s=h_\phi(s)$\Comment{Get representations}
\For{$t=T,\dots,1$}\Comment{Denoise for $T$ steps}
\State $\bm{\epsilon}\sim\mathcal{N}(\bm{0},\bm{I})$\Comment{Sample Gaussian noise}
\State $\hat{\bm{x}}_{t-1}=\frac{\delta_{t-1}}{\delta_{t}}\gamma_t\hat{\bm{x}}_t+\frac{\hat{\delta}_{t}}{\delta_{t}}(1-\beta_{t-1})g_\theta(\hat{\bm{x}}_t,t,\hat{\bm{x}}_T) + (\beta_{t-1}-\frac{\delta_{t-1}}{\delta_t}\gamma_t\beta_t)\hat{\bm{x}}_T+\sqrt{\tilde{\delta}_t}\bm{\epsilon}$\Comment{Apply reverse process}
\EndFor 
\State \Return $\hat{\bm{x}}_0$\Comment{Return the embedding for next item prediction}
\end{algorithmic}
\end{algorithm}

%% file: 5_exp.tex
\section{Experiment}

In this section, we conduct a series of experiments to answer the following research questions:

\noindent \textbf{RQ1}: How does BBDRec perform on real-world datasets compared to other baseline methods?

\noindent \textbf{RQ2}: What is the impact of the individual components of BBDRec on its effectiveness?

\noindent \textbf{RQ3}: How do the specific hyper-parameters of BBDRec affect its performance?

\noindent \textbf{RQ4}: What insights can be obtained from the learned representation distributions and diffusion trajectories of BBDRec?

\subsection{Experimental Setting}

\begin{table}[t]
\caption{Statistical details of the evaluation datasets, where AvgLen denotes the average historical sequence length per user.}
\label{tab:dataset}
\centering
\begin{tabular}{lcccccc}
\hline
Dataset & Baby & Beauty & ML-100K & Sports & Toys & Yelp \\ 
\hline
\#Users & 11,761 & 10,553 & 938 & 22,686 & 11,268 & 136,346 \\
\#Items & 4,731 & 6,086 & 1,008 & 12,301 & 7,309 & 64,669 \\
\#Interactions & 92,613 & 94,119 & 54,457 & 185,779 & 95,468 & 1,857,033 \\
AvgLen & 7.89 & 8.92 & 58.01 & 8.19 & 8.47 & 13.62 \\
Sparsity & 99.62\% & 99.74\% & 94.50\% & 99.63\% & 99.95\% & 99.98\% \\
\hline
\end{tabular}
\end{table}

\subsubsection{Datasets}
We conduct extensive experiments on six widely used public recommendation datasets, including four subsets from the Amazon Review dataset, namely \textbf{Baby}, \textbf{Beauty}, \textbf{Sports}, and \textbf{Toys}, together with \textbf{ML-100K} and \textbf{Yelp}. Following prior work~\cite{SASRec}, we employ the standard 5-core filtering strategy, retaining only users and items with at least five interactions. For evaluation, we adopt the widely used \textit{leave-one-out} protocol, where the most recent interaction of each user is used for testing, the second most recent interaction is used for validation, and the remaining interactions are used for training. Moreover, following prior studies~\cite{SASRec,DiffuRec}, the maximum length of user interaction sequences is truncated to 50. The summary statistics of the preprocessed datasets are presented in Table~\ref{tab:dataset}.

\subsubsection{Baselines}
We compare BBDRec with representative recurrent, variational, Transformer-based, frequency-enhanced, and diffusion-based sequential recommendation models as follows:
\begin{itemize}[leftmargin=*]
    \item \textbf{GRU4Rec}~\cite{GRU4Rec}. This method employs a recurrent neural network (GRU) for session-based recommendation.

    \item \textbf{SVAE}~\cite{SVAE}. This method extends variational autoencoders with a recurrent neural network to capture temporal dependencies in user interaction sequences for sequential recommendation.
    
    \item \textbf{SASRec}~\cite{SASRec}. This method leverages the causal Transformer to capture users' historical preferences.

    \item \textbf{BERT4Rec}~\cite{BERT4Rec}. This method incorporates bidirectional Transformer layers and adopts a Cloze-style objective to model user sequential patterns.

    \item \textbf{CORE}~\cite{CORE}. This method proposes a representation-consistent framework that unifies session and item embedding spaces through a linear encoder and robust distance metric for session-based recommendation.
    
    \item \textbf{LightSANs}~\cite{LightSANs}. This method introduces a low-rank decomposition into the self-attention mechanism, improving both the efficiency and effectiveness of Transformer-based sequential recommendation models.

    \item \textbf{FEARec}~\cite{FEARec}. This method enhances self-attention in the frequency domain to jointly capture high-frequency signals and periodic patterns in user behavior sequences for sequential recommendation.
    
    \item \textbf{EulerFormer}~\cite{EulerFormer}. This method proposes a unified theoretical framework to capture both semantic and positional differences between items within Transformers, thereby enhancing expressive power in sequential modeling.
    
    \item \textbf{DreamRec}~\cite{DreamRec}. This method employs a classifier-free guidance scheme~\cite{cfd_guidance}, utilizing historical sequence representations to guide the generation of oracle items.
    
    \item \textbf{DiffuRec}~\cite{DiffuRec}. This method utilizes Transformers as the denoising network and incorporates uncertainty injection into the diffusion process.
    
    \item \textbf{SdifRec}~\cite{SdifRec}. This method employs the Schrödinger bridge framework to address information loss during the transition process by replacing the Gaussian prior in the diffusion model with the user's current state.
\end{itemize}

\subsubsection{Implementation Details}
Overall, we follow the experimental settings in prior work for fair comparison~\cite{DiffuRec}. During evaluation, we rank the ground-truth item against all candidate items and report HR@20 and NDCG@20 under the full-ranking protocol. We conduct training for up to 500 epochs, and perform validation every 5 epochs to monitor model performance. An early stopping strategy is adopted with a patience of 4 validation rounds, terminating training when no improvement is observed on the validation set. For optimization, we use the Adam optimizer with a batch size of 512. The initial learning rate is set to 0.001, and a cosine annealing schedule is applied to adjust the learning rate during training. The embedding dimension and hidden size are both fixed to 128 across all models to ensure consistency and to make the overall number of parameters comparable across different methods. For discriminative sequential recommendation baselines, we use the standard cross-entropy objective over the item set when applicable. For diffusion-based baselines, we follow their original training objectives and use a truncated linear noise schedule~\cite{DiffuRec}, with the number of diffusion steps set to the default value of 32, consistent with prior work. Each method is evaluated over five independent runs, and we report the average performance for a robust comparison.

For our proposed BBDRec, the sequence encoder is implemented as a 2-layer Transformer, while the denoiser is also instantiated as a 2-layer Transformer by default (we additionally compare with an MLP-based variant in the experiments). For the embedding parameters in Stage 2, we set the learning rate to 0 for the first 5 epochs, and then switch to the default learning rate schedule for subsequent training. We apply a causal mask to enable next-token-prediction (NTP)-style training. The variance scale $m$ in Equation~\eqref{eq:BB} is tuned within $\{0.5, 1, 2, 4, 8\}$, and the diffusion step $T$ is selected from $\{2, 4, 8, 16, 32\}$. The loss weight $\eta$ is searched within $\{0.01, 0.1, 1, 10\}$, while the aggregation weight $\lambda$ in Equation~\eqref{eq:agg} is tuned from $\{0.001, 0.01, 0.1, 1\}$. 

\subsection{Overall Performance (RQ1)}
\begin{table*}[t]
\centering
\caption{Main results (\%) on six datasets. The best results are highlighted in bold, while the second-best results are underlined. RI denotes the relative improvement over the second-best result when BBDRec ranks first; a dash indicates that BBDRec does not achieve the best result.}
\label{tab:main}
\begin{tabular}{lcccccc}
\toprule
& \multicolumn{2}{c}{Baby} 
& \multicolumn{2}{c}{Beauty} 
& \multicolumn{2}{c}{ML-100K} \\
\cmidrule(lr){2-3} 
\cmidrule(lr){4-5} 
\cmidrule(lr){6-7}
Method 
& HR@20 & NDCG@20 
& HR@20 & NDCG@20 
& HR@20 & NDCG@20 \\
\midrule
GRU4Rec     
& 5.4576 & 2.2568 
& 12.6462 & 5.6347 
& 18.6858 & 7.1357 \\

SVAE        
& 2.7439 & 1.0086 
& 3.9204 & 1.4985 
& 8.1504 & 3.2905 \\

SASRec      
& 5.5268 & 2.5197 
& \underline{15.2038} & \underline{7.6509} 
& 18.4538 & 7.1034 \\

BERT4Rec    
& 4.0009 & 1.6198 
& 9.8723 & 3.9216 
& 10.1833 & 3.7504 \\

CORE        
& 2.7472 & 0.9427 
& 7.5635 & 2.6558 
& 11.8659 & 4.0904 \\

LightSANs   
& 4.0658 & 1.4990 
& 9.5874 & 4.5894 
& 14.1129 & 5.0443 \\

FEARec      
& 3.5786 & 1.5130 
& 9.7294 & 4.3869 
& 9.6959 & 3.5048 \\

EulerFormer 
& \underline{5.7906} & 2.4982 
& 14.7346 & 7.5428 
& \underline{19.3524} & \underline{7.6313} \\

DreamRec    
& 0.7648 & 0.2777 
& 0.6815 & 0.2728 
& 3.4395 & 1.3339 \\

DiffuRec    
& 5.3820 & \underline{2.5649} 
& 13.9102 & 7.3326 
& 16.0688 & 6.5550 \\

SdifRec     
& 2.8370 & 1.0849 
& 9.3768 & 4.9997 
& 13.9364 & 5.5393 \\

BBDRec      
& \textbf{7.4841} & \textbf{3.2566} 
& \textbf{17.1476} & \textbf{8.2508} 
& \textbf{22.9676} & \textbf{8.4818} \\

\rowcolor{gray!15}
\textit{RI} 
& \textit{+29.24\%} & \textit{+26.98\%} 
& \textit{+12.79\%} & \textit{+7.84\%} 
& \textit{+18.69\%} & \textit{+11.15\%} \\

\midrule[0.8pt]
& \multicolumn{2}{c}{Sports} 
& \multicolumn{2}{c}{Toys} 
& \multicolumn{2}{c}{Yelp} \\
\cmidrule(lr){2-3} 
\cmidrule(lr){4-5} 
\cmidrule(lr){6-7}
Method 
& HR@20 & NDCG@20 
& HR@20 & NDCG@20 
& HR@20 & NDCG@20 \\
\midrule
GRU4Rec     
& 6.0511 & 2.6069 
& 5.6216 & 2.5605 
& 6.3762 & 2.5408 \\

SVAE        
& 2.0019 & 0.8447 
& 1.5031 & 0.6266 
& 1.9238 & 0.7804 \\

SASRec      
& 6.4059 & \underline{3.3440} 
& 10.7082 & \underline{6.0517} 
& 6.6894 & 2.5277 \\

BERT4Rec    
& 4.2550 & 1.7844 
& 4.9423 & 1.9627 
& 4.0681 & 1.5224 \\

CORE        
& 2.8127 & 0.9802 
& 6.1481 & 2.2170 
& 2.3459 & 0.8788 \\

LightSANs   
& 4.4246 & 1.8463 
& 4.3054 & 1.8942 
& 3.7219 & 1.4837 \\

FEARec      
& 3.9261 & 1.6689 
& 6.2075 & 3.3492 
& 3.0050 & 1.0783 \\

EulerFormer 
& \underline{6.4759} & 3.2259 
& \underline{10.9120} & \textbf{6.2564} 
& 6.4350 & 2.4252 \\

DreamRec    
& 0.7432 & 0.2148 
& 0.4755 & 0.1731 
& 0.7681 & 0.2477 \\

DiffuRec    
& 6.2174 & 3.2067 
& 9.8590 & 5.8508 
& \underline{6.7315} & \underline{2.5951} \\

SdifRec     
& 1.9488 & 0.8677 
& 3.2694 & 1.7659 
& 1.0697 & 0.4083 \\

BBDRec      
& \textbf{7.9144} & \textbf{3.3872} 
& \textbf{12.1264} & 5.8464 
& \textbf{7.5349} & \textbf{3.0242} \\

\rowcolor{gray!15}
\textit{RI} 
& \textit{+22.21\%} & \textit{+1.29\%} 
& \textit{+11.13\%} & \textit{-} 
& \textit{+11.93\%} & \textit{+16.53\%} \\

\bottomrule
\end{tabular}
\end{table*}

We begin by assessing the overall recommendation performance of the compared methods. The summarized results are presented in Table~\ref{tab:main}, yielding the following observations:

\begin{itemize}[leftmargin=*]

\item \textbf{BBDRec achieves the best performance on most datasets and metrics, demonstrating its strong generalization ability.} This suggests that explicitly modeling the transition from user history to target items via a Brownian bridge process is more effective than relying on noise-anchored diffusion formulations, as it better preserves user-specific preference structures during generation.

\item \textbf{DreamRec shows relatively weak performance in our experimental setting.} 
One possible reason is that its diffusion-only supervision may provide insufficient item-level discriminative signals under the sequential recommendation setting, making it difficult to maintain a well-structured preference-aware embedding space. This further highlights the importance of our progressive learning strategy in stabilizing representation learning and enabling effective diffusion-based recommendation.

\item \textbf{SASRec and other Transformer-based methods are strong baselines and demonstrate competitive performance across multiple datasets.} Nevertheless, their purely discriminative nature may limit their ability to explicitly model stochastic preference transitions, suggesting that they can be further enhanced by integrating diffusion-based generative modeling.

\end{itemize}

\subsection{Ablation Study (RQ2)}

\begin{figure}[t]
    \centering
    \includegraphics[width=0.94\textwidth]{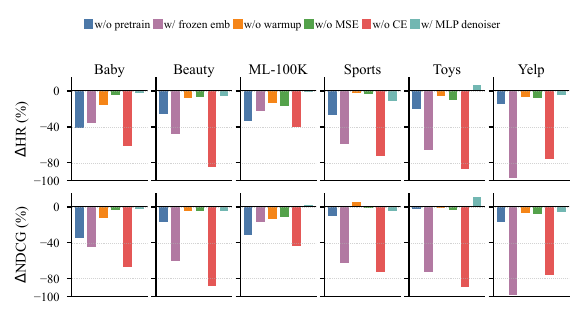}
    \caption{Ablation effects of key components in BBDRec on six datasets. The upper and lower rows report the relative changes in HR@20 and NDCG@20 compared with the base model, respectively.}
    \label{fig:abl}
    \Description{The figure shows the relative changes in HR@20 and NDCG@20 after removing or replacing key components of BBDRec, including pretraining, embedding tuning, warmup, MSE loss, CE loss, and the Transformer denoiser.}
\end{figure}

To substantiate the rationale behind different design choices in BBDRec, we conduct a comprehensive ablation study by systematically disabling or replacing one critical component at a time. Specifically, we introduce the following variants for comparison:

\begin{itemize}[leftmargin=*]

\item \textbf{w/o pretrain}. This variant removes the Stage-1 pretraining process in Section~\ref{sec:PLS} and directly trains all parameters in an end-to-end manner.

\item \textbf{w/ frozen emb}. This variant loads the pretrained embedding table from Stage-1 but keeps it frozen during Stage-2, so that only the remaining model parameters are updated in subsequent training.

\item \textbf{w/o warmup}. This variant removes the early-stage warm-up strategy in Stage-2, eliminating the controlled learning rate scheduling for sparse updates.

\item \textbf{w/o MSE}. This variant removes the MSE loss in Equation~\eqref{eq:BB-loss-diff} used for diffusion modeling.

\item \textbf{w/o CE}. This variant removes the cross-entropy loss in Equation~\eqref{eq:2_rec} used for recommendation learning.

\item \textbf{w/ MLP denoiser}. This variant replaces the Transformer-based denoiser $g_\theta(\cdot)$ with a simple MLP architecture.

\end{itemize}

Figure~\ref{fig:abl} illustrates the comparison results on six datasets, from which we draw the following observations:

\begin{itemize}[leftmargin=*]

\item \textbf{The pretraining stage is essential for effective optimization.}
Removing Stage-1 pretraining consistently degrades performance across datasets and metrics, showing that it provides a well-structured initialization for subsequent optimization. Meanwhile, freezing the pretrained embeddings during Stage-2 also causes clear performance drops, indicating that pretraining alone is insufficient and the embedding space must remain adaptable to diffusion-enhanced joint learning. Together, these results demonstrate the necessity of progressive optimization: Stage-1 captures collaborative and sequential patterns, while Stage-2 further refines the embeddings under joint recommendation and diffusion supervision. Without this design, the model either learns from scratch or fails to adapt pretrained representations to the preference-bridging diffusion process, leading to inferior representations.

\item \textbf{Warmup scheduling and the Transformer-based denoiser provide moderate improvements.} 
Removing the sparse-embedding warm-up strategy or replacing the Transformer-based denoiser with an MLP leads to relatively small performance drops, suggesting that these components are beneficial but not central to BBDRec. 
The warm-up strategy helps stabilize early optimization under sparse updates, while the Transformer denoiser offers stronger capacity for sequential denoising.

\item \textbf{The CE objective is critical for stable recommendation-oriented diffusion learning.} 
Removing the CE loss causes a much larger performance drop than removing the MSE loss, showing that item-level discriminative supervision is essential for BBDRec.
The CE objective provides item-level ranking signals that anchor embeddings to meaningful recommendation semantics and prevent representation collapse. 
In contrast, without the MSE loss, the model loses part of its diffusion-based denoising ability, but can still retain non-trivial recommendation capability through CE-supervised item-level ranking.

\end{itemize}

\begin{figure}[t]
    \centering
    \includegraphics[width=0.94\textwidth]{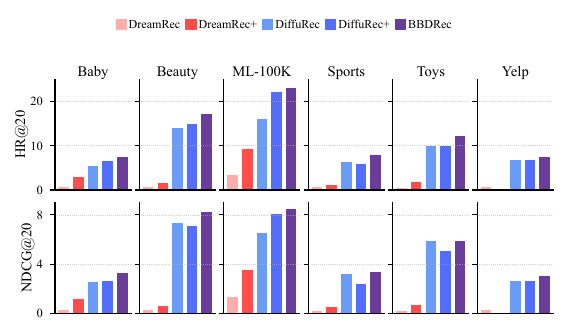}
    \caption{Effect of applying the proposed progressive learning strategy to existing diffusion-based sequential recommendation models. DreamRec+ and DiffuRec+ denote the variants enhanced with our learning strategy.}
    \label{fig:7_plus}
   \Description{The figure compares the original DreamRec and DiffuRec models with their enhanced variants using the proposed progressive learning strategy across multiple datasets and metrics.}
\end{figure}

Based on the above findings, we further examine whether the proposed progressive learning strategy can benefit existing diffusion-based sequential recommendation models. 
To this end, we apply the same learning strategy to DreamRec and DiffuRec, which are representative diffusion-based baselines for sequential recommendation, obtaining DreamRec+ and DiffuRec+, respectively. As shown in Figure~\ref{fig:7_plus}, the proposed strategy improves these models in several cases, but the gains are not consistent across all datasets and metrics. 
Moreover, although DreamRec+ and DiffuRec+ outperform their original counterparts in some cases, they still lag behind BBDRec. 
These results indicate that stable training alone is insufficient for effective diffusion-based recommendation; a recommendation-aligned transition process, such as the Brownian bridge formulation in BBDRec, is also crucial.

\subsection{Hyperparameter Analysis (RQ3)}

\begin{figure}[t]
    \centering
    \includegraphics[width=0.94\textwidth]{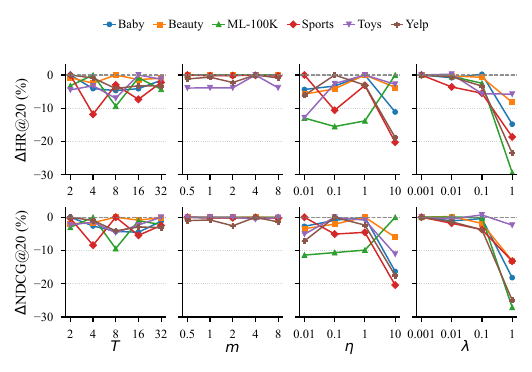}
    \caption{Hyperparameter sensitivity of BBDRec with respect to $T$, $m$, $\eta$, and $\lambda$. The upper and lower rows report the relative changes in HR@20 and NDCG@20, respectively.}
    \label{fig:hyper}
    \Description{The figure reports the relative changes in HR@20 and NDCG@20 when varying the diffusion step, variance scale, loss weight, and aggregation weight of BBDRec.}
\end{figure}

In our investigation, several key hyper-parameters are involved in controlling the behavior of BBDRec. 
Specifically, the variance scale $m$ in Equation~\eqref{eq:BB} determines the stochastic intensity of the Brownian bridge diffusion process, while the diffusion step $T$ specifies the number of intermediate transitions used for representation transformation. 
In addition, the loss weight $\eta$ balances the diffusion-oriented denoising objective and the recommendation-oriented ranking objective, and $\lambda$ in Equation~\eqref{eq:agg} controls the weight for aggregating different input components. 
To evaluate their impact, we conduct a systematic sensitivity analysis by varying one hyper-parameter at a time while keeping the others fixed. 
Specifically, $m$ is searched within $\{0.5, 1, 2, 4, 8\}$, $T$ is selected from $\{2, 4, 8, 16, 32\}$, $\eta$ is tuned within $\{0.01, 0.1, 1, 10\}$, and $\lambda$ is selected from $\{0.001, 0.01, 0.1, 1\}$. The relative performance changes are summarized in Figure~\ref{fig:hyper}, from which we draw the following observations:

\begin{itemize}[leftmargin=*]

\item \textbf{BBDRec requires only a small number of diffusion steps for effective refinement.}
Although varying $T$ leads to moderate performance fluctuations, competitive results can already be achieved with relatively few diffusion steps. This is because BBDRec performs diffusion-based refinement on behavior-aware representations instead of reconstructing from pure noise, making a small $T$ sufficient in most cases.
This property suggests the potential efficiency advantage of BBDRec over noise-anchored diffusion methods that typically require more sampling steps.

\item \textbf{BBDRec is robust to the variance scale $m$.}
The performance curves with respect to $m$ remain relatively flat on most datasets. This indicates that the variance scale mainly affects the smoothness of the Brownian bridge transition, while BBDRec can maintain stable recommendation performance across a broad range of stochastic scales.

\item \textbf{The loss weight $\eta$ is comparatively more sensitive.}
The curves show that inappropriate values of $\eta$, especially overly large ones, can lead to clear performance degradation. This is because $\eta$ balances the diffusion-oriented MSE loss and the recommendation-oriented CE loss. An improper balance may either weaken the discriminative ranking supervision or overemphasize the denoising objective, thereby impairing the final recommendation quality.

\item \textbf{The aggregation weight $\lambda$ should remain relatively small.} The performance curves show that small $\lambda$ values yield stable results, whereas an overly large $\lambda$ leads to clear degradation. This indicates that timestep embeddings are useful for providing diffusion-step information, but should not dominate the aggregated representation. Otherwise, they may overwhelm behavior-aware semantic signals and hurt recommendation performance.

\end{itemize}

\subsection{In-Depth Analysis (RQ4)}

To further understand the internal mechanism and optimization behavior of BBDRec, we conduct an in-depth analysis from two perspectives: trajectory evolution and embedding distribution. The trajectory analysis focuses on how representations evolve during the forward process, while the embedding analysis investigates the structural properties of the learned embedding space.

\subsubsection{Trajectory Analysis}
\begin{figure}[t]
    \centering
    \includegraphics[width=0.74\textwidth]{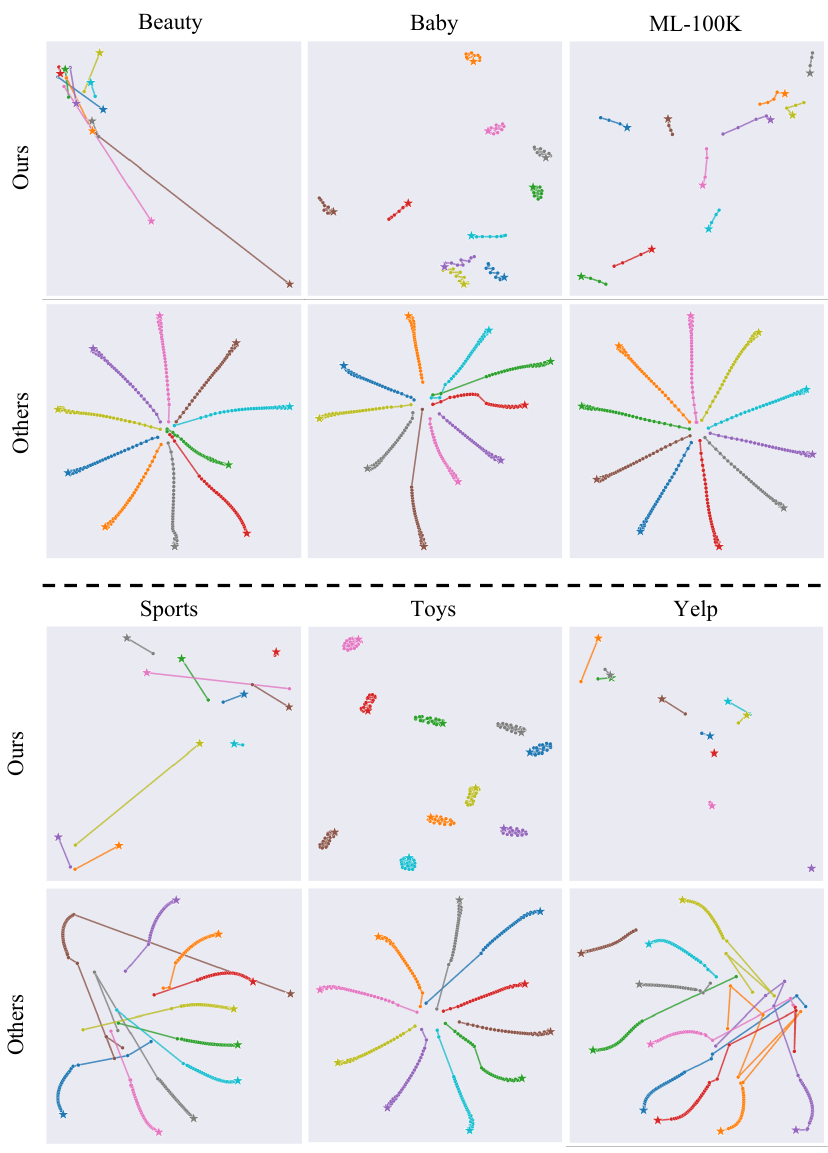}
    \caption{t-SNE visualization of diffusion trajectories on six datasets. Rows correspond to different diffusion paradigms, and columns correspond to datasets. BBDRec follows preference-centric trajectories, while the baseline follows noise-anchored trajectories. The star marker denotes the representation of the target item.}
    \label{fig:traj}
    \Description{The figure visualizes forward diffusion trajectories of sampled instances using t-SNE across six datasets. It compares noise-anchored diffusion methods with BBDRec, showing that BBDRec produces more compact trajectories between item and history representations.}
\end{figure}

We compare the mechanism of BBDRec, which follows a preference-centric diffusion paradigm, with DreamRec and DiffuRec, which are representative noise-anchored diffusion methods. 
To this end, we randomly sample 10 instances and visualize their noising trajectories during the forward process. 

As observed in Figure~\ref{fig:traj}, the trajectories of noise-anchored methods are generally dispersed and exhibit a radial pattern, where the target item representations are progressively perturbed toward noisy states with limited semantic structure during the forward process.
Although such a noise-anchored formulation provides a standard diffusion mechanism, it gradually destroys the semantic information of the target representation and drives intermediate states away from the behavior-aware preference space.
As a result, the forward trajectories become widely scattered, making the reverse process rely on long-range reconstruction from noisy representations.

In contrast, BBDRec presents more compact and preference-centric trajectories.
Starting from the target item representation, BBDRec leverages the Brownian bridge formulation to guide the forward process toward behavior-aware user history representations, so that the noising trajectory is constrained between two semantically meaningful endpoints instead of drifting toward unstructured noise.
Consequently, the intermediate states remain more semantically aligned with user preferences, and the reverse process can be viewed as preference-oriented refinement instead of full reconstruction from noise.
In this sense, the user history acts as a \emph{semantic compass}, guiding the reverse process toward the target item rather than leaving it to navigate from unstructured noise.

\subsubsection{Embedding Analysis}

\begin{figure}[]
    \centering
    \includegraphics[width=0.84\textwidth]{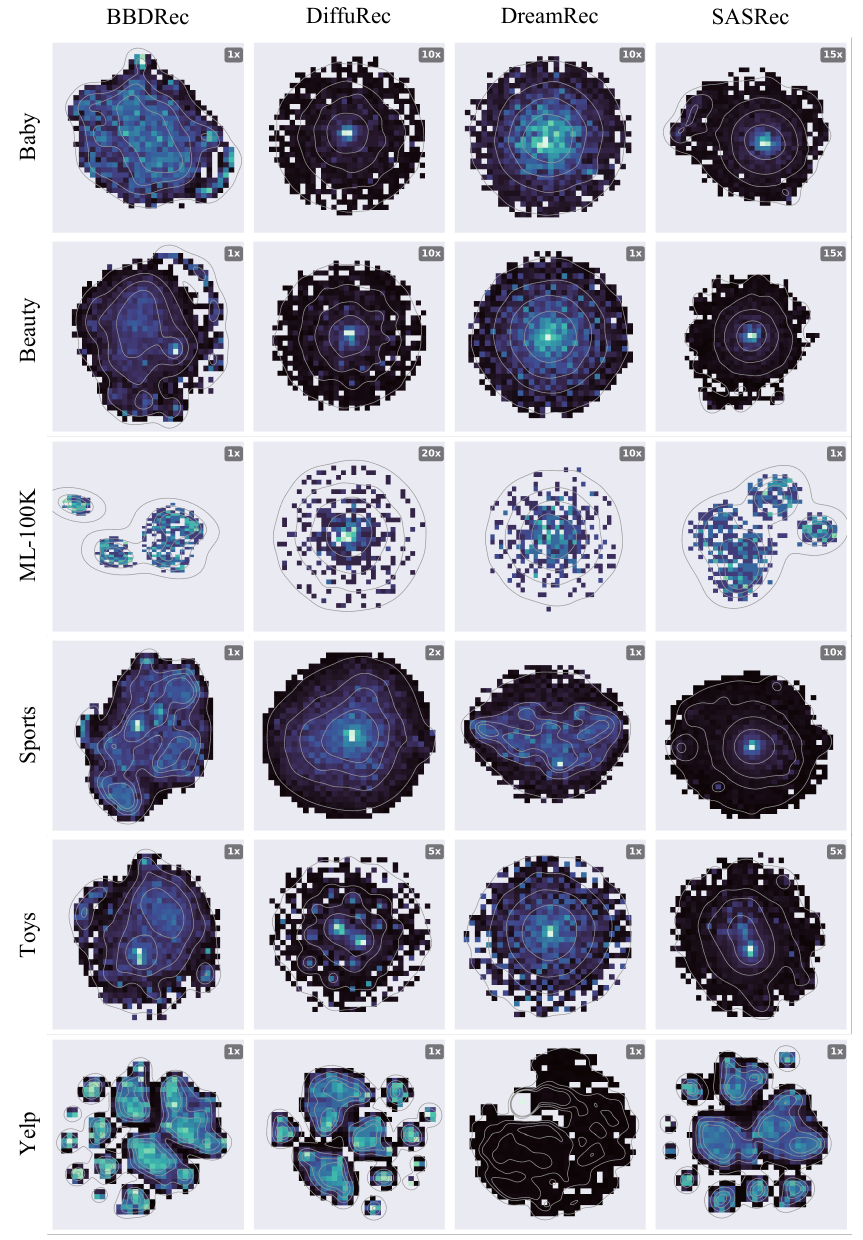}
    \caption{t-SNE visualization of embedding distributions across six datasets. Rows correspond to different datasets, and columns correspond to different methods. The zoom level is indicated in the top-right corner.}
    \label{fig:emb}
    \Description{The figure shows t-SNE projections and density contours of learned item embeddings for SASRec, DiffuRec, DreamRec, and BBDRec across six datasets, illustrating differences in embedding distribution and structural organization.}
\end{figure}

We further investigate the structural properties of the learned embedding space. 
Specifically, we compare BBDRec with three representative baselines, including SASRec, DiffuRec, and DreamRec. 
For each method, we extract the learned embedding table after training and apply t-SNE to project the high-dimensional embeddings into a two-dimensional space. 
To better illustrate the distributional structure, we visualize both the projected embedding points and their density contours. 
The results are shown in Figure~\ref{fig:emb}, where rows correspond to datasets and columns correspond to different methods. 
The value in the upper-right corner of each subfigure indicates the magnification factor used for visualization.

As observed, different methods lead to clearly different embedding distributions. 
SASRec, as a purely discriminative sequential recommendation model, can learn representations with certain discriminative capability. 
DiffuRec and DreamRec introduce diffusion-based modeling, which may indicate a tendency toward over-smoothed or less discriminative representation distributions.
This suggests that noise-anchored diffusion may impose a Gaussian-prior-driven smoothing effect on the embedding space, which may hinder the preservation of fine-grained preference semantics and reduce the expressiveness of item representations.

In contrast, BBDRec generally forms more structured and semantically organized embedding distributions. 
Across most datasets, its embeddings exhibit clearer high-density regions and more coherent local neighborhoods, while still preserving sufficient distributional diversity. 
This can be attributed to two aspects. 
First, by establishing a Brownian bridge transition between target item representations and behavior-aware user history representations, BBDRec weakens the Gaussian-distribution regularization imposed by conventional noise-anchored diffusion, making the diffusion objective more consistent with the recommendation objective. 
Second, the progressive learning strategy provides a more informative and stable initialization for subsequent joint optimization, which further facilitates the formation of a compact yet expressive embedding space.

These observations further point to a broader perspective on the role of diffusion in generative recommendation~\cite{kuaishou_survey}. 
Beyond the next-token prediction (NTP) paradigm, diffusion offers an alternative way to model recommendation generation: instead of predicting discrete item IDs or semantic tokens, it enables generation by sampling next-item representations in a continuous embedding space. 
This opens up a potential alternative route beyond ID-token-based generative recommenders such as TIGER~\cite{TIGER}. From this perspective, our analysis highlights two key requirements for continuous-space generative recommendation. 
The first is to construct an item embedding space with sufficient discriminability, so that generated representations can faithfully preserve fine-grained item semantics. 
The second is to define a semantically meaningful transition process, so that generation is not merely a noise-to-data reconstruction problem, but a preference-aligned evolution toward the target item. 
Addressing these two challenges is essential for making diffusion a principled and effective paradigm for generative recommendation.

%% file: 6_rel.tex
\section{Related Work}
In this section, we review existing research on sequential recommendation and diffusion-based recommendation.

\subsection{Sequential Recommendation}
Sequential recommendation leverages users' historical interactions to predict the next item they are likely to engage with. Early studies predominantly relied on Markov chains to model item transition patterns~\cite{FuseSim,FPMC}. Subsequent research explored a variety of deep learning-based architectures~\cite{GRU4Rec,caser,SASRec,BERT4Rec}. For example, GRU4Rec~\cite{GRU4Rec} introduced recurrent neural networks (RNNs) into session-based recommendation, while Caser~\cite{caser} employed convolutional neural networks (CNNs) to capture sequential patterns.
SASRec~\cite{SASRec} and BERT4Rec~\cite{BERT4Rec} further leveraged self-attention mechanisms to capture long-range dependencies in user behavior sequences.
To further incorporate rich contextual information, FDSA~\cite{FDSA} integrates item attributes into self-attention blocks to enhance item representations. Similarly, $\text{S}^3$-Rec~\cite{S3Rec} maximizes mutual information across different forms of contextual data to improve sequential recommendation. Despite their strong performance, these methods are mostly formulated as discriminative models that directly estimate item-level preference scores, which may limit their ability to explicitly model the generative process of user preference evolution.

More recently, generative recommendation has attracted increasing attention for its potential to model recommendation as a generation problem~\cite{GR, BIGRec, LLM4GR, li2024survey,kuaishou_survey}. Generally, these models either directly utilize item IDs/text embeddings or tokenize them into semantic identifiers, thereby formulating recommendation as a sequence-to-sequence generation task.
For instance, TIGER~\cite{TIGER} creates semantically meaningful codeword tuples as semantic IDs for items and autoregressively decodes the identifiers of target candidates.
BIGRec~\cite{BIGRec} proposes a bi-step grounding paradigm, where large language models (LLMs) are first fine-tuned to generate meaningful item-related tokens, which are then grounded to actual items.
HSTU~\cite{HSTU} reformulates recommendation problems as sequential transduction tasks within a generative modeling framework and introduces a new architecture tailored for high-cardinality, non-stationary streaming recommendation data. 
StreamingVQ~\cite{StreamingVQ} presents a real-time indexing retrieval model that effectively captures semantic changes in industry recommender systems, offering additional benefits such as index balancing and repairability.

\subsection{Diffusion Models for Recommendation}
Diffusion models have emerged as a powerful class of deep generative models and have achieved remarkable success in applications such as image generation and inpainting~\cite{DiffSurvey1, DiffSurvey2}. Recent studies have explored their use in recommender systems~\cite{Diff4RecSurvey}, which can be broadly categorized into two directions. The first direction applies diffusion models to generate or reconstruct user-item interaction distributions over the entire item set~\cite{DiffRec, CODIGEM, CDDRec, DiffuASR, PDRec, LD4MRec}.
For example, DiffRec~\cite{DiffRec} reduces the noise scale and shortens the diffusion process to better preserve personalized information in user interaction vectors.
DiffuASR~\cite{DiffuASR} introduces a diffusion-based augmentation framework for sequential recommendation, generating high-quality pseudo sequences to address data sparsity and the long-tail user problem. 
PDRec~\cite{PDRec} introduces a plug-in diffusion model to infer users' dynamic preferences over all items, alleviating data sparsity through positive augmentation and improving model optimization with noise-free negative sampling.
However, these approaches face scalability challenges in real-world scenarios with a large number of items.

To address this scalability issue, the second direction adapts diffusion models to directly generate item embeddings. Within this line of research, most studies formulate next-item prediction as a noise-anchored generation process~\cite{DreamRec, DiffuRec, DimeRec, CaDiRec, DiffRIS, iDreamRec, SdifRec}, drawing inspiration from text-guided image generation frameworks~\cite{selfguidance, cfd_guidance}.
Specifically, these methods incorporate the representation of sequential history as additional input to the denoising model, guiding the generation of the next item's embedding. For instance, DreamRec~\cite{DreamRec} adopts a classifier-free guidance scheme~\cite{cfd_guidance}, leveraging historical sequence representations encoded by SASRec~\cite{SASRec} to steer the generation of target item embeddings. 
CaDiRec~\cite{CaDiRec} leverages a context-aware diffusion model to generate semantically consistent augmented views for contrastive learning, enhancing sequential recommendation by addressing the limitations of random augmentation. 
DimeRec~\cite{DimeRec} utilizes ComiRec~\cite{ComiRec} for guidance extraction to generate the next user interest representation and mitigate the instability caused by non-stationary user histories. SdifRec~\cite{SdifRec} employs the Schrödinger bridge framework to mitigate information loss within the diffusion process while maintaining a bridge-based stochastic formulation. Nevertheless, Schrödinger bridge methods often require iterative approximation or simulation procedures, which may introduce additional computational complexity~\cite{DDBM}. In contrast, BBDRec uses a linear Brownian bridge to place the diffusion trajectory between two meaningful semantic anchors: the target item representation and the user history representation. This design provides a simple and tractable transition path, better aligning the denoising objective with recommendation-oriented preference modeling.

%% file: 7_con.tex
\section{Conclusion}

This paper revisited sequential recommendation from a preference-centric diffusion perspective and identified a key limitation of existing diffusion-based methods: they typically follow a history-guided denoising paradigm built on an ``item $\leftrightarrow$ noise'' formulation, where item representations are reconstructed from Gaussian noise conditioned on user history. This design may introduce a detour through the noise space and weaken the explicit modeling of the semantic relationship between target items and user history. To address this issue, we proposed BBDRec, which uses a linear Brownian bridge to place the diffusion trajectory between two meaningful semantic anchors: target item representations and user history representations. This simple yet principled ``item $\leftrightarrow$ history'' formulation provides a closed-form preference-centric transition path, enabling a more direct utilization of historical information and making diffusion learning more consistent with recommendation-oriented optimization. Together with a progressive learning strategy, BBDRec further improves training stability and recommendation effectiveness.

Despite its effectiveness, BBDRec still relies on sufficient user historical interactions, which may limit its performance in cold-start scenarios. In future work, we will explore incorporating multimodal item features and other side information to improve robustness under sparse interactions. We also plan to evaluate the proposed framework in large-scale real-world systems to further validate its effectiveness and scalability.

%% file: 8_appd.tex
\appendix

\section{Theoretical Derivation of BBDRec}\label{sec:app}

\subsection{Forward Process (Proof of Equation~\eqref{eq:BB-forward})}\label{sec:dev-forward}

From Equation~\eqref{eq:BB}, given $\bm{x}_0$ and $\bm{x}_T$, the intermediate state can be written as:
\begin{equation}\label{eq:app-x-t}
\bm{x}_t
=
(1-\beta_t)\bm{x}_0
+
\beta_t\bm{x}_T
+
\sqrt{\delta_t}\bm{\epsilon}_t,
\end{equation}
and
\begin{equation}\label{eq:app-x-t-1}
\bm{x}_{t-1}
=
(1-\beta_{t-1})\bm{x}_0
+
\beta_{t-1}\bm{x}_T
+
\sqrt{\delta_{t-1}}\bm{\epsilon}_{t-1},
\end{equation}
where $\beta_t=\frac{t}{T}$ and $\delta_t=m\cdot\beta_t(1-\beta_t)$. From Equation~\eqref{eq:app-x-t-1}, we have:
\begin{equation}
\bm{x}_0
=
\frac{
\bm{x}_{t-1}
-
\beta_{t-1}\bm{x}_T
-
\sqrt{\delta_{t-1}}\bm{\epsilon}_{t-1}
}{
1-\beta_{t-1}
}.
\end{equation}
Substituting it into Equation~\eqref{eq:app-x-t}, we obtain:
\begin{align}
\bm{x}_t
&=
\gamma_t\bm{x}_{t-1}
+
(\beta_t-\gamma_t\beta_{t-1})\bm{x}_T
+
\sqrt{\delta_t}\bm{\epsilon}_t
-
\gamma_t\sqrt{\delta_{t-1}}\bm{\epsilon}_{t-1},
\label{eq:app-forward-sub}
\end{align}
where
\begin{equation}
\gamma_t=\frac{1-\beta_t}{1-\beta_{t-1}}.
\end{equation}
Since the Brownian bridge is a Gaussian process, any linear combination of its states is also Gaussian. Therefore, the conditional transition can be written as:
\begin{equation}\label{eq:temp_}
q(\bm{x}_t|\bm{x}_{t-1},\bm{x}_T)
=
\mathcal{N}
\left(
\bm{x}_t;
\gamma_t\bm{x}_{t-1}
+
(\beta_t-\gamma_t\beta_{t-1})\bm{x}_T,
\hat{\delta}_{t}\bm{I}
\right).
\end{equation}

Next, we derive the variance term $\hat{\delta}_{t}$. Note that
$\sqrt{\frac{T}{m}\delta_t}\bm{\epsilon}_t$ and
$\sqrt{\frac{T}{m}\delta_{t-1}}\bm{\epsilon}_{t-1}$ can be interpreted as
$B(t)$ and $B(t-1)$, respectively, where $B(\cdot)$ denotes the standard Brownian bridge~\cite{BB}. For any $t_i,t_j\in\{0,\cdots,T\}$, it satisfies
\begin{equation}
\mathrm{Cov}\left(B(t_i),B(t_j)\right)
=
\min(t_i,t_j)-\frac{t_it_j}{T}.
\end{equation}
Then $\hat{\delta}_t$ is the variance of the stochastic term in Equation~\eqref{eq:app-forward-sub}, \emph{i.e.},
\begin{align}
\hat{\delta}_{t}
&=
\mathrm{Var}
\left(
\sqrt{\delta_t}\bm{\epsilon}_{t}
-
\gamma_t\sqrt{\delta_{t-1}}\bm{\epsilon}_{t-1}
\right)
\notag \\
&=
\delta_t
+
\gamma_t^2\delta_{t-1}
-
2\gamma_t
\mathrm{Cov}
\left(
\sqrt{\delta_t}\bm{\epsilon}_{t},
\sqrt{\delta_{t-1}}\bm{\epsilon}_{t-1}
\right)
\notag \\
&=
\delta_t
+
\gamma_t^2\delta_{t-1}
-
2\gamma_t
\frac{m}{T}
\left(
t-1-\frac{t(t-1)}{T}
\right)
\notag \\
&=
\delta_t
+
\gamma_t^2\delta_{t-1}
-
2\gamma_t
\frac{m(T-t)(t-1)}{T^2}
\notag \\
&=
\delta_t
-
\gamma_t^2\delta_{t-1}.
\label{eq:app-forward-var}
\end{align}
This is consistent with the form in Equation~\eqref{eq:BB-forward}.

\subsection{Reverse Process (Proof of Equation~\eqref{eq:BB-reverse})}\label{sec:ded-reverse}

According to Bayes' theorem, we have
\begin{equation}
q(\bm{x}_{t-1}|\bm{x}_t,\bm{x}_0,\bm{x}_T)
=
\frac{
q(\bm{x}_t|\bm{x}_{t-1},\bm{x}_0,\bm{x}_T)
q(\bm{x}_{t-1}|\bm{x}_0,\bm{x}_T)
}{
q(\bm{x}_t|\bm{x}_0,\bm{x}_T)
}.
\end{equation}
By the Markov property of the forward process,
$q(\bm{x}_t|\bm{x}_{t-1},\bm{x}_0,\bm{x}_T)
=
q(\bm{x}_t|\bm{x}_{t-1},\bm{x}_T)$, whose form has been derived in Equation~\eqref{eq:temp_}. In addition, $q(\bm{x}_{t-1}|\bm{x}_0,\bm{x}_T)$ and $q(\bm{x}_t|\bm{x}_0,\bm{x}_T)$ are given by Equation~\eqref{eq:BB}.

For notational simplicity, we present the derivation in the scalar form; the vector case follows by applying the same derivation independently to each dimension. Substituting the Gaussian densities gives
\begin{align}
&q(\bm{x}_{t-1}|\bm{x}_t,\bm{x}_0,\bm{x}_T)
\notag \\
&=
\frac{
\mathcal{N}
\left(
\bm{x}_t;
\gamma_t\bm{x}_{t-1}
+
(\beta_t-\gamma_t\beta_{t-1})\bm{x}_T,
\hat{\delta}_{t}\bm{I}
\right)
\mathcal{N}
\left(
\bm{x}_{t-1};
(1-\beta_{t-1})\bm{x}_0+\beta_{t-1}\bm{x}_T,
\delta_{t-1}\bm{I}
\right)
}{
\mathcal{N}
\left(
\bm{x}_t;
(1-\beta_t)\bm{x}_0+\beta_t\bm{x}_T,
\delta_t\bm{I}
\right)
}.
\label{eq:app-reverse-gaussian}
\end{align}
Since the denominator does not depend on $\bm{x}_{t-1}$, we only keep terms involving $\bm{x}_{t-1}$:
\begin{align}
&q(\bm{x}_{t-1}|\bm{x}_t,\bm{x}_0,\bm{x}_T)
\notag \\
&\propto
\exp
\left\{
-\frac{1}{2}
\left[
\frac{
\left(
\bm{x}_t
-
\gamma_t\bm{x}_{t-1}
-
(\beta_t-\gamma_t\beta_{t-1})\bm{x}_T
\right)^2
}{
\hat{\delta}_{t}
}
+
\frac{
\left(
\bm{x}_{t-1}
-
(1-\beta_{t-1})\bm{x}_0
-
\beta_{t-1}\bm{x}_T
\right)^2
}{
\delta_{t-1}
}
\right]
\right\}.
\label{eq:app-reverse-exp}
\end{align}
Collecting the quadratic and linear terms of $\bm{x}_{t-1}$, we obtain
\begin{align}
q(\bm{x}_{t-1}|\bm{x}_t,\bm{x}_0,\bm{x}_T)
&\propto
\exp
\left\{
-\frac{1}{2}
\left(
U\bm{x}_{t-1}^2
+
2V\bm{x}_{t-1}
\right)
\right\}
\notag \\
&\propto
\exp
\left\{
-\frac{
\left(
\bm{x}_{t-1}
+
\frac{V}{U}
\right)^2
}{
2/U
}
\right\}
\notag \\
&=
\mathcal{N}
\left(
\bm{x}_{t-1};
-\frac{V}{U},
\frac{1}{U}\bm{I}
\right).
\label{eq:app-reverse-square}
\end{align}
Here $U$ and $V$ are introduced for notational convenience. Specifically,
\begin{align}
U
&=
\frac{\gamma_t^2}{\hat{\delta}_{t}}
+
\frac{1}{\delta_{t-1}}
\notag \\
&=
\frac{\hat{\delta}_{t}+\gamma_t^2\delta_{t-1}}
{\hat{\delta}_{t}\delta_{t-1}}
=
\frac{\delta_t}{\hat{\delta}_{t}\delta_{t-1}},
\label{eq:app-U}
\end{align}
where the last equality follows from
$\hat{\delta}_{t}=\delta_t-\gamma_t^2\delta_{t-1}$. Similarly, the linear coefficient is
\begin{align}
V
&=
\frac{
-\gamma_t\bm{x}_t
+
\gamma_t(\beta_t-\gamma_t\beta_{t-1})\bm{x}_T
}{
\hat{\delta}_{t}
}
-
\frac{
(1-\beta_{t-1})\bm{x}_0
+
\beta_{t-1}\bm{x}_T
}{
\delta_{t-1}
}
\notag \\
&=
-\frac{\gamma_t}{\hat{\delta}_{t}}\bm{x}_t
+
\frac{\beta_{t-1}-1}{\delta_{t-1}}\bm{x}_0
+
\frac{
\gamma_t\beta_t\delta_{t-1}
-
\beta_{t-1}
\left(
\gamma_t^2\delta_{t-1}
+
\hat{\delta}_{t}
\right)
}{
\hat{\delta}_{t}\delta_{t-1}
}
\bm{x}_T
\notag \\
&=
-\frac{\gamma_t}{\hat{\delta}_{t}}\bm{x}_t
+
\frac{\beta_{t-1}-1}{\delta_{t-1}}\bm{x}_0
+
\frac{
\gamma_t\beta_t\delta_{t-1}
-
\beta_{t-1}\delta_t
}{
\hat{\delta}_{t}\delta_{t-1}
}
\bm{x}_T.
\label{eq:app-V}
\end{align}
Therefore, the posterior mean and variance are
\begin{equation}
\bm{\mu}_t(\bm{x}_t,\bm{x}_0,\bm{x}_T)
=
-\frac{V}{U},
\qquad
\tilde{\delta}_t
=
\frac{1}{U}.
\end{equation}
It is straightforward to verify that these results are consistent with the form in Equation~\eqref{eq:BB-reverse}.

\subsection{Training Objective (Proof of Equation~\eqref{eq:BB-loss-diff})}\label{sec:ded-training}

According to the definition of the evidence lower bound (ELBO), we have
\begin{align}
\mathrm{ELBO}
&=
\mathbb{E}_{q(\bm{x}_{1:T}|\bm{x}_0,\bm{x}_T)}
\left[
\log
\frac{
p(\bm{x}_{0:T}|\bm{x}_T)
}{
q(\bm{x}_{1:T}|\bm{x}_0,\bm{x}_T)
}
\right]
\notag \\
&=
\mathbb{E}_{q(\bm{x}_{1:T}|\bm{x}_0,\bm{x}_T)}
\left[
\log
\frac{
p(\bm{x}_T|\bm{x}_T)
\prod_{t=1}^{T}
p_\theta(\bm{x}_{t-1}|\bm{x}_{t},\bm{x}_T)
}{
\prod_{t=1}^{T}
q(\bm{x}_t|\bm{x}_{t-1},\bm{x}_T)
}
\right].
\label{eq:app-elbo-def}
\end{align}
By separating the reconstruction term and the intermediate denoising terms, we obtain
\begin{align}
\mathrm{ELBO}
&=
\mathbb{E}_{q(\bm{x}_{1:T}|\bm{x}_0,\bm{x}_T)}
\left[
\log
\frac{
p(\bm{x}_T|\bm{x}_T)
p_\theta(\bm{x}_0|\bm{x}_1,\bm{x}_T)
}{
q(\bm{x}_T|\bm{x}_{T-1},\bm{x}_T)
}
\right] +
\sum_{t=1}^{T-1}
\mathbb{E}_{q(\bm{x}_{1:T}|\bm{x}_0,\bm{x}_T)}
\left[
\log
\frac{
p_\theta(\bm{x}_t|\bm{x}_{t+1},\bm{x}_T)
}{
q(\bm{x}_t|\bm{x}_{t-1},\bm{x}_T)
}
\right].
\label{eq:app-elbo-middle}
\end{align}
Following the standard variational decomposition for diffusion models, this can be rewritten as
\begin{align}
\mathrm{ELBO}
&=
\underbrace{
\mathbb{E}_{q(\bm{x}_1|\bm{x}_0,\bm{x}_T)}
\left[
\log p_\theta(\bm{x}_0|\bm{x}_1,\bm{x}_T)
\right]
}_{\text{reconstruction term}}
-
\sum_{t=2}^{T}
\underbrace{
\mathbb{E}_{q(\bm{x}_t|\bm{x}_0,\bm{x}_T)}
\left[
D_{\mathrm{KL}}
\left(
q(\bm{x}_{t-1}|\bm{x}_t,\bm{x}_0,\bm{x}_T)
\Vert
p_\theta(\bm{x}_{t-1}|\bm{x}_t,\bm{x}_T)
\right)
\right]
}_{\text{denoising matching term}}.
\label{eq:ELBO}
\end{align}
Here, the ELBO consists of a reconstruction term and multiple denoising matching terms.

The denoising matching term encourages the modeling distribution
$p_\theta(\bm{x}_{t-1}|\bm{x}_t,\bm{x}_T)$ to approximate the tractable posterior
$q(\bm{x}_{t-1}|\bm{x}_t,\bm{x}_0,\bm{x}_T)$ through KL divergence. Since the variances of the two Gaussian distributions are set to match exactly, optimizing the KL divergence reduces to minimizing the distance between their means. By employing the reparameterization method defined in Equation~\eqref{eq:BB-reparam}, the denoising matching term can be computed as
\begin{align}
\mathcal{L}^t
&\triangleq
\mathbb{E}_{q(\bm{x}_t|\bm{x}_0,\bm{x}_T)}
\left[
D_{\mathrm{KL}}
\left(
q(\bm{x}_{t-1}|\bm{x}_t,\bm{x}_0,\bm{x}_T)
\Vert
p_\theta(\bm{x}_{t-1}|\bm{x}_t,\bm{x}_T)
\right)
\right]
\notag \\
&=
\mathbb{E}_{q(\bm{x}_t|\bm{x}_0,\bm{x}_T)}
\left[
\frac{1}{2\tilde{\delta}_t}
\left\|
\bm{\mu}_t(\bm{x}_t,\bm{x}_0,\bm{x}_T)
-
\hat{\bm{\mu}}_\theta(\bm{x}_t,t,\bm{x}_T)
\right\|_2^2
\right]
\notag \\
&=
\mathbb{E}_{q(\bm{x}_t|\bm{x}_0,\bm{x}_T)}
\left[
\frac{1}{2\tilde{\delta}_t}
\left(
\frac{\hat{\delta}_t}{\delta_t}
(1-\beta_{t-1})
\right)^2
\left\|
\bm{x}_0
-
g_\theta(\bm{x}_t,t,\bm{x}_T)
\right\|_2^2
\right].
\label{eq:Lt}
\end{align}

For the reconstruction term, we define
\begin{align}
\mathcal{L}^1
&\triangleq
-
\mathbb{E}_{q(\bm{x}_1|\bm{x}_0,\bm{x}_T)}
\left[
\log p_\theta(\bm{x}_0|\bm{x}_1,\bm{x}_T)
\right]
\notag \\
&\approx
\mathbb{E}_{q(\bm{x}_1|\bm{x}_0,\bm{x}_T)}
\left[
\left\|
\bm{x}_0
-
g_\theta(\bm{x}_1,1,\bm{x}_T)
\right\|_2^2
\right],
\label{eq:L1}
\end{align}
where the Gaussian log-likelihood
$\log p_\theta(\bm{x}_0|\bm{x}_1,\bm{x}_T)$ is approximated by the unweighted negative squared error
$-\|\bm{x}_0-g_\theta(\bm{x}_1,1,\bm{x}_T)\|_2^2$, as discussed in~\cite{VAE4CF}.

According to Equation~\eqref{eq:Lt} and Equation~\eqref{eq:L1}, maximizing the ELBO in Equation~\eqref{eq:ELBO} is equivalent to minimizing
$\sum_{t=1}^{T}\mathcal{L}^{t}.
$
Following prior work~\cite{DiffRec}, we omit the time-dependent weighting term and optimize the simplified denoising objective. In practical implementation, we uniformly sample a timestep $t$ to optimize $\mathcal{L}^t$, resulting in the expectation form of Equation~\eqref{eq:BB-loss-diff}.

%% file: 9_ref.bib
@inproceedings{DreamRec,
author = {Yang, Zhengyi and Wu, Jiancan and Wang, Zhicai and Wang, Xiang and Yuan, Yancheng and He, Xiangnan},
booktitle = {Advances in Neural Information Processing Systems},
editor = {A. Oh and T. Naumann and A. Globerson and K. Saenko and M. Hardt and S. Levine},
pages = {24247--24261},
publisher = {Curran Associates, Inc.},
title = {Generate What You Prefer: Reshaping Sequential Recommendation via Guided Diffusion},
volume = {36},
year = {2023}
}

@article{DiffuRec,
author = {Li, Zihao and Sun, Aixin and Li, Chenliang},
title = {DiffuRec: A Diffusion Model for Sequential Recommendation},
year = {2023},
issue_date = {May 2024},
publisher = {Association for Computing Machinery},
address = {New York, NY, USA},
volume = {42},
number = {3},
issn = {1046-8188},
journal = {ACM Trans. Inf. Syst.},
month = dec,
articleno = {66},
numpages = {28}
}

@inproceedings{SdifRec,
author = {Xie, Wenjia and Zhou, Rui and Wang, Hao and Shen, Tingjia and Chen, Enhong},
title = {Bridging User Dynamics: Transforming Sequential Recommendations with Schr\"{o}dinger Bridge and Diffusion Models},
year = {2024},
isbn = {9798400704369},
publisher = {Association for Computing Machinery},
address = {New York, NY, USA},
booktitle = {Proceedings of the 33rd ACM International Conference on Information and Knowledge Management},
pages = {2618–2628},
numpages = {11},
location = {Boise, ID, USA},
series = {CIKM '24}
}

@inproceedings{DimeRec,
author = {Li, Wuchao and Huang, Rui and Zhao, Haijun and Liu, Chi and Zheng, Kai and Liu, Qi and Mou, Na and Zhou, Guorui and Lian, Defu and Song, Yang and Bao, Wentian and Yu, Enyun and Ou, Wenwu},
title = {DimeRec: A Unified Framework for Enhanced Sequential Recommendation via Generative Diffusion Models},
year = {2025},
isbn = {9798400713293},
publisher = {Association for Computing Machinery},
address = {New York, NY, USA},
booktitle = {Proceedings of the Eighteenth ACM International Conference on Web Search and Data Mining},
pages = {726–734},
numpages = {9},
location = {Hannover, Germany},
series = {WSDM '25}
}

@INPROCEEDINGS{SASRec,
author={Kang, Wang-Cheng and McAuley, Julian},
booktitle={2018 IEEE International Conference on Data Mining (ICDM)}, 
title={Self-Attentive Sequential Recommendation}, 
year={2018},
volume={},
number={},
pages={197-206},
}

@inproceedings{caser,
author = {Tang, Jiaxi and Wang, Ke},
title = {Personalized Top-N Sequential Recommendation via Convolutional Sequence Embedding},
year = {2018},
isbn = {9781450355810},
publisher = {Association for Computing Machinery},
address = {New York, NY, USA},
booktitle = {Proceedings of the Eleventh ACM International Conference on Web Search and Data Mining},
pages = {565–573},
numpages = {9},
location = {Marina Del Rey, CA, USA},
series = {WSDM '18}
}

@inproceedings{GRU4Rec,
author = {Hidasi, Bal\'{a}zs and Karatzoglou, Alexandros},
title = {Recurrent Neural Networks with Top-k Gains for Session-based Recommendations},
year = {2018},
isbn = {9781450360142},
publisher = {Association for Computing Machinery},
address = {New York, NY, USA},
booktitle = {Proceedings of the 27th ACM International Conference on Information and Knowledge Management},
pages = {843–852},
numpages = {10},
location = {Torino, Italy},
series = {CIKM '18}
}

@inproceedings{DiffRec,
author = {Wang, Wenjie and Xu, Yiyan and Feng, Fuli and Lin, Xinyu and He, Xiangnan and Chua, Tat-Seng},
title = {Diffusion Recommender Model},
year = {2023},
isbn = {9781450394086},
publisher = {Association for Computing Machinery},
address = {New York, NY, USA},
booktitle = {Proceedings of the 46th International ACM SIGIR Conference on Research and Development in Information Retrieval},
pages = {832–841},
numpages = {10},
location = {Taipei, Taiwan},
series = {SIGIR '23}
}

@inproceedings{CDDRec,
author = {Wang, Yu and Liu, Zhiwei and Yang, Liangwei and Yu, Philip S.},
title = {Conditional Denoising Diffusion for Sequential Recommendation},
year = {2024},
isbn = {978-981-97-2264-8},
publisher = {Springer-Verlag},
address = {Berlin, Heidelberg},
booktitle = {Advances in Knowledge Discovery and Data Mining: 28th Pacific-Asia Conference on Knowledge Discovery and Data Mining, PAKDD 2024, Taipei, Taiwan, May 7–10, 2024, Proceedings, Part V},
pages = {156–169},
numpages = {14},
location = {Taipei, Taiwan}
}

@inproceedings{DiffuASR,
author = {Liu, Qidong and Yan, Fan and Zhao, Xiangyu and Du, Zhaocheng and Guo, Huifeng and Tang, Ruiming and Tian, Feng},
title = {Diffusion Augmentation for Sequential Recommendation},
year = {2023},
isbn = {9798400701245},
publisher = {Association for Computing Machinery},
address = {New York, NY, USA},
booktitle = {Proceedings of the 32nd ACM International Conference on Information and Knowledge Management},
pages = {1576–1586},
numpages = {11},
location = {Birmingham, United Kingdom},
series = {CIKM '23}
}

@article{PDRec, 
title={Plug-In Diffusion Model for Sequential Recommendation}, 
volume={38}, 
number={8}, 
journal={Proceedings of the AAAI Conference on Artificial Intelligence}, author={Ma, Haokai and Xie, Ruobing and Meng, Lei and Chen, Xin and Zhang, Xu and Lin, Leyu and Kang, Zhanhui}, 
year={2024}, 
month={Mar.}, 
pages={8886-8894} 
}

@inproceedings{CaDiRec,
author = {Cui, Ziqiang and Wu, Haolun and He, Bowei and Cheng, Ji and Ma, Chen},
title = {Context Matters: Enhancing Sequential Recommendation with Context-aware Diffusion-based Contrastive Learning},
year = {2024},
isbn = {9798400704369},
publisher = {Association for Computing Machinery},
address = {New York, NY, USA},
booktitle = {Proceedings of the 33rd ACM International Conference on Information and Knowledge Management},
pages = {404–414},
numpages = {11},
location = {Boise, ID, USA},
series = {CIKM '24}
}

@article{iDreamRec,
title={Generate and Instantiate What You Prefer: Text-Guided Diffusion for Sequential Recommendation},
author={Hu, Guoqing and Yang, Zhengyi and Cai, Zhibo and Zhang, An and Wang, Xiang},
journal={arXiv preprint arXiv:2410.13428},
year={2024},
}

@inproceedings{DiffRIS,
author = {Niu, Yong and Xing, Xing and Jia, Zhichun and Liu, Ruidi and Xin, Mindong and Cui, Jianfu},
title = {Diffusion Recommendation with Implicit Sequence Influence},
year = {2024},
isbn = {9798400701726},
publisher = {Association for Computing Machinery},
address = {New York, NY, USA},
booktitle = {Companion Proceedings of the ACM Web Conference 2024},
pages = {1719–1725},
numpages = {7},
location = {Singapore, Singapore},
series = {WWW '24}
}

@article{LD4MRec,
author = {Zhu, Jiarui and Hou, Jun and Yu, Penghang and Tan, Zhiyi and Bao, Bing-Kun},
title = {LD4MRec: simplifying and powering diffusion model for multimedia recommendation},
year = {2025},
issue_date = {Aug 2025},
publisher = {Springer-Verlag},
address = {Berlin, Heidelberg},
volume = {31},
number = {5},
issn = {0942-4962},
journal = {Multimedia Syst.},
month = aug,
numpages = {13}
}

@inproceedings{BERT4Rec,
author = {Sun, Fei and Liu, Jun and Wu, Jian and Pei, Changhua and Lin, Xiao and Ou, Wenwu and Jiang, Peng},
title = {BERT4Rec: Sequential Recommendation with Bidirectional Encoder Representations from Transformer},
year = {2019},
isbn = {9781450369763},
publisher = {Association for Computing Machinery},
address = {New York, NY, USA},
booktitle = {Proceedings of the 28th ACM International Conference on Information and Knowledge Management},
pages = {1441–1450},
numpages = {10},
location = {Beijing, China},
series = {CIKM '19}
}

@inproceedings{HSTU,
author = {Zhai, Jiaqi and Liao, Lucy and Liu, Xing and Wang, Yueming and Li, Rui and Cao, Xuan and Gao, Leon and Gong, Zhaojie and Gu, Fangda and He, Jiayuan and Lu, Yinghai and Shi, Yu},
title = {Actions speak louder than words: trillion-parameter sequential transducers for generative recommendations},
year = {2025},
publisher = {JMLR.org},
booktitle = {Proceedings of the 41st International Conference on Machine Learning},
articleno = {2414},
numpages = {26},
location = {Vienna, Austria},
series = {ICML'24}
}

@inproceedings{TIGER,
author = {Rajput, Shashank and Mehta, Nikhil and Singh, Anima and Keshavan, Raghunandan and Vu, Trung and Heidt, Lukasz and Hong, Lichan and Tay, Yi and Tran, Vinh Q. and Samost, Jonah and Kula, Maciej and Chi, Ed H. and Sathiamoorthy, Maheswaran},
title = {Recommender systems with generative retrieval},
year = {2024},
publisher = {Curran Associates Inc.},
address = {Red Hook, NY, USA},
booktitle = {Proceedings of the 37th International Conference on Neural Information Processing Systems},
articleno = {452},
numpages = {17},
location = {New Orleans, LA, USA},
series = {NIPS '23},
}

@article{BIGRec,
author = {Bao, Keqin and Zhang, Jizhi and Wang, Wenjie and Zhang, Yang and Yang, Zhengyi and Luo, Yanchen and Chen, Chong and Feng, Fuli and Tian, Qi},
title = {A Bi-Step Grounding Paradigm for Large Language Models in Recommendation Systems},
year = {2025},
issue_date = {December 2025},
publisher = {Association for Computing Machinery},
address = {New York, NY, USA},
volume = {3},
number = {4},
journal = {ACM Trans. Recomm. Syst.},
month = apr,
articleno = {53},
numpages = {27}
}

@inproceedings{GR,
author = {Wang, Wenjie and Lin, Xinyu and Feng, Fuli and He, Xiangnan and Chua, Tat-Seng},
title = {Generative Recommendation: Towards Personalized Multimodal Content Generation},
year = {2025},
isbn = {9798400713316},
publisher = {Association for Computing Machinery},
address = {New York, NY, USA},
booktitle = {Companion Proceedings of the ACM on Web Conference 2025},
pages = {2421–2425},
numpages = {5},
location = {Sydney NSW, Australia},
series = {WWW '25}
}

@inproceedings{DDPM,
author = {Ho, Jonathan and Jain, Ajay and Abbeel, Pieter},
title = {Denoising diffusion probabilistic models},
year = {2020},
isbn = {9781713829546},
publisher = {Curran Associates Inc.},
address = {Red Hook, NY, USA},
booktitle = {Proceedings of the 34th International Conference on Neural Information Processing Systems},
articleno = {574},
numpages = {12},
location = {Vancouver, BC, Canada},
series = {NIPS '20},
}

@article{Diff4RecSurvey,
title={A Survey on Diffusion Models for Recommender Systems},
author={Lin, Jianghao and Liu, Jiaqi and Zhu, Jiachen and Xi, Yunjia and Liu, Chengkai and Zhang, Yangtian and Yu, Yong and Zhang, Weinan},
journal={arXiv preprint arXiv:2409.05033},
year={2024},
}

@inproceedings{selfguidance,
author = {Epstein, Dave and Jabri, Allan and Poole, Ben and Efros, Alexei and Holynski, Aleksander},
booktitle = {Advances in Neural Information Processing Systems},
editor = {A. Oh and T. Naumann and A. Globerson and K. Saenko and M. Hardt and S. Levine},
pages = {16222--16239},
publisher = {Curran Associates, Inc.},
title = {Diffusion Self-Guidance for Controllable Image Generation},
volume = {36},
year = {2023}
}

@inproceedings{cfd_guidance,
title={Classifier-Free Diffusion Guidance},
author={Ho, Jonathan and Salimans, Tim},
booktitle={NeurIPS 2021 Workshop on Deep Generative Models and Downstream Applications},
year={2021},
}

@inproceedings{BBDM,
author={Li, Bo and Xue, Kaitao and Liu, Bin and Lai, Yu-Kun},
booktitle={2023 IEEE/CVF Conference on Computer Vision and Pattern Recognition (CVPR)}, 
title={BBDM: Image-to-Image Translation with Brownian Bridge Diffusion Models}, 
year={2023},
volume={},
number={},
pages={1952-1961},
}

@inproceedings{SBBB,
author = {Wang, Peiyong and Xiao, Bohan and He, Qisheng and Glide-Hurst, Carri and Dong, Ming},
title = {Score-Based Image-to-Image Brownian Bridge},
year = {2024},
isbn = {9798400706868},
publisher = {Association for Computing Machinery},
address = {New York, NY, USA},
booktitle = {Proceedings of the 32nd ACM International Conference on Multimedia},
pages = {10765–10773},
numpages = {9},
location = {Melbourne VIC, Australia},
series = {MM '24}
}

@ARTICLE{DiffSurvey1,
author={Croitoru, Florinel-Alin and Hondru, Vlad and Ionescu, Radu Tudor and Shah, Mubarak},
journal={IEEE Transactions on Pattern Analysis and Machine Intelligence}, 
title={Diffusion Models in Vision: A Survey}, 
year={2023},
volume={45},
number={9},
pages={10850-10869},
}

@article{DiffSurvey2,
author = {Yang, Ling and Zhang, Zhilong and Song, Yang and Hong, Shenda and Xu, Runsheng and Zhao, Yue and Zhang, Wentao and Cui, Bin and Yang, Ming-Hsuan},
title = {Diffusion Models: A Comprehensive Survey of Methods and Applications},
year = {2023},
issue_date = {April 2024},
publisher = {Association for Computing Machinery},
address = {New York, NY, USA},
volume = {56},
number = {4},
issn = {0360-0300},
journal = {ACM Comput. Surv.},
month = nov,
articleno = {105},
numpages = {39}
}

@article{Understanding,
title={Understanding diffusion models: A unified perspective},
author={Luo, Calvin},
journal={arXiv preprint arXiv:2208.11970},
year={2022},
}

@InProceedings{LDM,
author    = {Rombach, Robin and Blattmann, Andreas and Lorenz, Dominik and Esser, Patrick and Ommer, Bj\"orn},
title     = {High-Resolution Image Synthesis With Latent Diffusion Models},
booktitle = {Proceedings of the IEEE/CVF Conference on Computer Vision and Pattern Recognition (CVPR)},
month     = {June},
year      = {2022},
pages     = {10684-10695},
numpages = {12},
}

@inproceedings{LLM4GR,
title = "Large Language Models for Generative Recommendation: A Survey and Visionary Discussions",
author = "Li, Lei  and
  Zhang, Yongfeng  and
  Liu, Dugang  and
  Chen, Li",
booktitle = "Proceedings of the 2024 Joint International Conference on Computational Linguistics, Language Resources and Evaluation (LREC-COLING 2024)",
month = may,
year = "2024",
address = "Torino, Italia",
publisher = "ELRA and ICCL",
pages = "10146--10159",
}

@article{li2024survey,
title={A survey of generative search and recommendation in the era of large language models},
author={Li, Yongqi and Lin, Xinyu and Wang, Wenjie and Feng, Fuli and Pang, Liang and Li, Wenjie and Nie, Liqiang and He, Xiangnan and Chua, Tat-Seng},
journal={arXiv preprint arXiv:2404.16924},
year={2024},
}

@inproceedings{CODIGEM,
title={Recommendation via collaborative diffusion generative model},
author={Walker, Joojo and Zhong, Ting and Zhang, Fengli and Gao, Qiang and Zhou, Fan},
booktitle={International Conference on Knowledge Science, Engineering and Management},
pages={593--605},
year={2022},
organization={Springer},
}

@inproceedings{ComiRec,
author = {Cen, Yukuo and Zhang, Jianwei and Zou, Xu and Zhou, Chang and Yang, Hongxia and Tang, Jie},
title = {Controllable Multi-Interest Framework for Recommendation},
year = {2020},
isbn = {9781450379984},
publisher = {Association for Computing Machinery},
address = {New York, NY, USA},
booktitle = {Proceedings of the 26th ACM SIGKDD International Conference on Knowledge Discovery \& Data Mining},
pages = {2942–2951},
numpages = {10},
location = {Virtual Event, CA, USA},
series = {KDD '20}
}

@article{BB,
title={Brownian bridge},
author={Chow, Winston C},
journal={Wiley interdisciplinary reviews: computational statistics},
volume={1},
number={3},
pages={325--332},
year={2009},
publisher={Wiley Online Library},
}

@inproceedings{VAE4CF,
author = {Liang, Dawen and Krishnan, Rahul G. and Hoffman, Matthew D. and Jebara, Tony},
title = {Variational Autoencoders for Collaborative Filtering},
year = {2018},
isbn = {9781450356398},
publisher = {International World Wide Web Conferences Steering Committee},
address = {Republic and Canton of Geneva, CHE},
booktitle = {Proceedings of the 2018 World Wide Web Conference},
pages = {689–698},
numpages = {10},
location = {Lyon, France},
series = {WWW '18}
}

@inproceedings{DDBM,
title={Denoising Diffusion Bridge Models},
author={Linqi Zhou and Aaron Lou and Samar Khanna and Stefano Ermon},
booktitle={The Twelfth International Conference on Learning Representations},
publisher    = {OpenReview.net},
year={2024},
}

@article{guidance_scale,
title={Bridging the Gap: Addressing Discrepancies in Diffusion Model Training for Classifier-Free Guidance},
author={Patel, Niket and Salamanca, Luis and Barba, Luis},
journal={arXiv preprint arXiv:2311.00938},
year={2023},
}

@InProceedings{SCFG,
author    = {Shen, Dazhong and Song, Guanglu and Xue, Zeyue and Wang, Fu-Yun and Liu, Yu},
title     = {Rethinking the Spatial Inconsistency in Classifier-Free Diffusion Guidance},
booktitle = {Proceedings of the IEEE/CVF Conference on Computer Vision and Pattern Recognition (CVPR)},
month     = {June},
year      = {2024},
pages     = {9370-9379},
}

@article{NoiseRefine,
title={A Noise is Worth Diffusion Guidance},
author={Ahn, Donghoon and Kang, Jiwon and Lee, Sanghyun and Min, Jaewon and Kim, Minjae and Jang, Wooseok and Cho, Hyoungwon and Paul, Sayak and Kim, SeonHwa and Cha, Eunju and others},
journal={arXiv preprint arXiv:2412.03895},
year={2024},
}

@inproceedings{FDSA,
author = {Zhang, Tingting and Zhao, Pengpeng and Liu, Yanchi and Sheng, Victor S. and Xu, Jiajie and Wang, Deqing and Liu, Guanfeng and Zhou, Xiaofang},
title = {Feature-level deeper self-attention network for sequential recommendation},
year = {2019},
isbn = {9780999241141},
publisher = {AAAI Press},
booktitle = {Proceedings of the 28th International Joint Conference on Artificial Intelligence},
pages = {4320–4326},
numpages = {7},
location = {Macao, China},
series = {IJCAI'19},
}

@inproceedings{S3Rec,
author = {Zhou, Kun and Wang, Hui and Zhao, Wayne Xin and Zhu, Yutao and Wang, Sirui and Zhang, Fuzheng and Wang, Zhongyuan and Wen, Ji-Rong},
title = {S3-Rec: Self-Supervised Learning for Sequential Recommendation with Mutual Information Maximization},
year = {2020},
isbn = {9781450368599},
publisher = {Association for Computing Machinery},
address = {New York, NY, USA},
booktitle = {Proceedings of the 29th ACM International Conference on Information \& Knowledge Management},
pages = {1893–1902},
numpages = {10},
location = {Virtual Event, Ireland},
series = {CIKM '20}
}

@inproceedings{FPMC,
author = {Rendle, Steffen and Freudenthaler, Christoph and Schmidt-Thieme, Lars},
title = {Factorizing personalized Markov chains for next-basket recommendation},
year = {2010},
isbn = {9781605587998},
publisher = {Association for Computing Machinery},
address = {New York, NY, USA},
booktitle = {Proceedings of the 19th International Conference on World Wide Web},
pages = {811–820},
numpages = {10},
location = {Raleigh, North Carolina, USA},
series = {WWW '10}
}

@INPROCEEDINGS{FuseSim,
author={He, Ruining and McAuley, Julian},
booktitle={2016 IEEE 16th International Conference on Data Mining (ICDM)}, 
title={Fusing Similarity Models with Markov Chains for Sparse Sequential Recommendation}, 
year={2016},
volume={},
number={},
pages={191-200},
}

@inproceedings{StreamingVQ,
author = {Bin, Xingyan and Cui, Jianfei and Yan, Wujie and Zhao, Zhichen and Han, Xintian and Yan, Chongyang and Zhang, Feng and Zhou, Xun and Yang, Xiao and Liu, Zuotao},
title = {Real-time Indexing for Large-scale Recommendation by Streaming Vector Quantization Retriever},
year = {2025},
isbn = {9798400714542},
publisher = {Association for Computing Machinery},
address = {New York, NY, USA},
booktitle = {Proceedings of the 31st ACM SIGKDD Conference on Knowledge Discovery and Data Mining V.2},
pages = {4273–4283},
numpages = {11},
location = {Toronto ON, Canada},
series = {KDD '25}
}

@article{kuaishou_survey,
year = {2025},
month = {December},
publisher = {Preprints},
author = {Xiaopeng Li and Bo Chen and Junda She and Shiteng Cao and You Wang and Qinlin Jia and Haiying He and Zheli Zhou and Zhao Liu and Ji Liu and Zhiyang Zhang and Yu Zhou and Guoping Tang and Yiqing Yang and Chengcheng Guo and Si Dong and Kuo Cai and Pengyue Jia and Maolin Wang and Wanyu Wang and Shiyao Wang and Xinchen Luo and Qigen Hu and Qiang Luo and Xiao Lv and Chaoyi Ma and Ruiming Tang and Kun Gai and Guorui Zhou and Xiangyu Zhao},
title = {A Survey of Generative Recommendation from a Tri-Decoupled Perspective: Tokenization, Architecture, and Optimization},
journal = {Preprints}
}

@inproceedings{embedding_collapse,
author = {Guo, Xingzhuo and Pan, Junwei and Wang, Ximei and Chen, Baixu and Jiang, Jie and Long, Mingsheng},
title = {On the embedding collapse when scaling up recommendation models},
year = {2024},
publisher = {JMLR.org},
booktitle = {Proceedings of the 41st International Conference on Machine Learning},
articleno = {671},
numpages = {19},
location = {Vienna, Austria},
series = {ICML'24}
}

@inproceedings{PreferDiff,
author       = {Shuo Liu and
              An Zhang and
              Guoqing Hu and
              Hong Qian and
              Tat{-}Seng Chua},
title        = {Preference Diffusion for Recommendation},
booktitle    = {The Thirteenth International Conference on Learning Representations,
              {ICLR} 2025, Singapore, April 24-28, 2025},
publisher    = {OpenReview.net},
year         = {2025},
}

@article{multi-epoch,
title={Multi-Epoch Learning for Deep Click-Through Rate Prediction Models},
author={Liu, Zhaocheng and Fan, Zhongxiang and Liang, Jian and Kong, Dongying and Li, Han},
journal={arXiv preprint arXiv:2305.19531},
year={2023}
}

@inproceedings{LightSANs,
author = {Fan, Xinyan and Liu, Zheng and Lian, Jianxun and Zhao, Wayne Xin and Xie, Xing and Wen, Ji-Rong},
title = {Lighter and Better: Low-Rank Decomposed Self-Attention Networks for Next-Item Recommendation},
year = {2021},
isbn = {9781450380379},
publisher = {Association for Computing Machinery},
address = {New York, NY, USA},
booktitle = {Proceedings of the 44th International ACM SIGIR Conference on Research and Development in Information Retrieval},
pages = {1733–1737},
numpages = {5},
location = {Virtual Event, Canada},
series = {SIGIR '21}
}

@inproceedings{EulerFormer,
author = {Tian, Zhen and Zhao, Wayne Xin and Zhang, Changwang and Zhao, Xin and Ma, Zhongrui and Wen, Ji-Rong},
title = {EulerFormer: Sequential User Behavior Modeling with Complex Vector Attention},
year = {2024},
isbn = {9798400704314},
publisher = {Association for Computing Machinery},
address = {New York, NY, USA},
booktitle = {Proceedings of the 47th International ACM SIGIR Conference on Research and Development in Information Retrieval},
pages = {1619–1628},
numpages = {10},
location = {Washington DC, USA},
series = {SIGIR '24}
}

@inproceedings{FEARec,
author = {Du, Xinyu and Yuan, Huanhuan and Zhao, Pengpeng and Qu, Jianfeng and Zhuang, Fuzhen and Liu, Guanfeng and Liu, Yanchi and Sheng, Victor S.},
title = {Frequency Enhanced Hybrid Attention Network for Sequential Recommendation},
year = {2023},
isbn = {9781450394086},
publisher = {Association for Computing Machinery},
address = {New York, NY, USA},
booktitle = {Proceedings of the 46th International ACM SIGIR Conference on Research and Development in Information Retrieval},
pages = {78–88},
numpages = {11},
location = {Taipei, Taiwan},
series = {SIGIR '23}
}

@inproceedings{SVAE,
author = {Sachdeva, Noveen and Manco, Giuseppe and Ritacco, Ettore and Pudi, Vikram},
title = {Sequential Variational Autoencoders for Collaborative Filtering},
year = {2019},
isbn = {9781450359405},
publisher = {Association for Computing Machinery},
address = {New York, NY, USA},
booktitle = {Proceedings of the Twelfth ACM International Conference on Web Search and Data Mining},
pages = {600–608},
numpages = {9},
location = {Melbourne VIC, Australia},
series = {WSDM '19}
}

@inproceedings{CORE,
author = {Hou, Yupeng and Hu, Binbin and Zhang, Zhiqiang and Zhao, Wayne Xin},
title = {CORE: Simple and Effective Session-based Recommendation within Consistent Representation Space},
year = {2022},
isbn = {9781450387323},
publisher = {Association for Computing Machinery},
address = {New York, NY, USA},
booktitle = {Proceedings of the 45th International ACM SIGIR Conference on Research and Development in Information Retrieval},
pages = {1796–1801},
numpages = {6},
location = {Madrid, Spain},
series = {SIGIR '22}
}

@article{CoLLM,
author = {Zhang, Yang and Feng, Fuli and Zhang, Jizhi and Bao, Keqin and Wang, Qifan and He, Xiangnan},
title = {CoLLM: Integrating Collaborative Embeddings Into Large Language Models for Recommendation},
year = {2025},
issue_date = {May 2025},
publisher = {IEEE Educational Activities Department},
address = {USA},
volume = {37},
number = {5},
issn = {1041-4347},
journal = {IEEE Trans. on Knowl. and Data Eng.},
month = may,
pages = {2329–2340},
numpages = {12}
}
